\documentclass[twocolumn]{aastex63}
\usepackage{times}
\usepackage{amsmath}
\usepackage{graphicx}
\usepackage{epstopdf}
\usepackage{hyperref}
\usepackage{threeparttable}

\newcommand{\beq}{\begin{equation}}
\newcommand{\eeq}{\end{equation}}
\newcommand{\bea}{\begin{eqnarray}}
\newcommand{\eea}{\end{eqnarray}}

\begin{document}

\title{\bf The UV/optical peak and X-ray brightening in TDE candidate AT2019azh: A case of stream-stream collision and delayed accretion} 

\author{Xiao-Long Liu}\thanks{liuxlong8@mail2.sysu.edu.cn}
\affiliation{School of Physics and Astronomy, Sun Yat-Sen University, Zhuhai, 519082, China}
\affiliation{CSST Science Center for the Guangdong-Hongkong-Macau Greater Bay Area, Sun Yat-Sen University, Zhuhai, 519082, China}
\author{Li-Ming Dou}
\affiliation{Center for Astrophysics, Guangzhou University, Guangzhou 510006, China}
\affiliation{Astronomy Science and Technology Research Laboratory of Department of Education of 
 Guangdong Province, Guangzhou 510006, China}

\author{Jin-Hong Chen}
\affiliation{School of Physics and Astronomy, Sun Yat-Sen University, Zhuhai, 519082, China}
\affiliation{CSST Science Center for the Guangdong-Hongkong-Macau Greater Bay Area, Sun Yat-Sen University, Zhuhai, 519082, China}
\author{Rong-Feng Shen}\thanks{shenrf3@mail.sysu.edu.cn}
\affiliation{School of Physics and Astronomy, Sun Yat-Sen University, Zhuhai, 519082, China}
\affiliation{CSST Science Center for the Guangdong-Hongkong-Macau Greater Bay Area, Sun Yat-Sen University, Zhuhai, 519082, China}
\begin{abstract}

We present and analyze the optical/UV and X-ray observations of a nearby tidal disruption event (TDE) candidate AT2019azh, spanning from $\sim$ 30 d before to $\sim$ 400 d after its early optical peak. The X-rays show a late brightening by a factor of $\sim$ 30-100 around 200 days after discovery, while the UV/opticals continuously decayed. The early X-rays show two flaring episodes of variation, temporally uncorrelated with the early UV/opticals. We found a clear sign of X-ray hardness evolution, i.e., the source is harder at early times, and becomes softer as it brightens later. The drastically different temporal behaviors in X-rays and UV/opticals suggest that the two bands are physically distinct emission components, and probably arise from different locations. These properties argue against the reprocessing of X-rays by any outflow as the origin of the UV/optical peak. The full data are best explained by a two-process scenario, in which the UV/optical peak is produced by the debris stream-stream collisions during the circularization phase; some shocked gas with low angular momentum forms an early, low-mass 'precursor' accretion disk which emits the early X-rays. The major body of the disk is formed after the circularization finishes, whose enhanced accretion rate produces the late X-ray brightening. AT2019azh is a strong case of TDE whose emission signatures of stream-stream collision and delayed accretion are both identified.

\end{abstract}


\section{Introduction}

In a tidal disruption event (TDE), a star wanders toward the center of a galactic nucleus and is tidally disrupted by a supermassive black hole (SMBH). A bright, multi-wavelength flare is produced when the stellar debris falls back and accretes toward the black hole. Observing such incidents is an important way to probe the otherwise dormant SMBHs in the center of many non-active galaxies and to study the feeding process of SMBHs. 

An unsettled issue still remains regarding the physical origin of the observed emission (i.e., where the emission is produced). Initially it was expected that the emission is produced in the accretion disk, which should peak in soft X-rays \citep{Rees1988,ulmer99}. However, most of the observed TDEs are X-ray weak\footnote{Note that a separate class of TDEs are non-thermal X-ray dominated, and they are thought to originate from a relativistic jet and require special orientation in order to be seen \citep{giannios11,burrows11,bloom11,levan11,zauderer11,cenko12,brown15}.}, only a small fraction of TDE sample have been detected with X-ray emission in follow-up observations: GALEX D1-9 and D3-13 \citep{Gezari08}, ASASSN-14li \citep{holoien16b}, AT2018fyk \citep{Wevers19} and ASASSN-15oi \citep{holoien16a}, with upper limits in a few cases: PS1-10jh \citep{Gezari12}, iPTF16fnl \citep{Blagorodnova17,Brown18}, and iPTF16axa \citep{Hung17}, while the majority were discovered in optical or UV \citep[e.g.,][]{Gezari12,Chornock2014,holoien14,van VelzenFarrar2014,Arcavi2014}. 

Two categories of models are proposed to explain the optical emissions. One involves ejected mass which reprocesses the high energy emission from the center to lower energies \citep{Ulmer1997,StrubbeQuataert2009,metzger16}. The other proposes on the stream-stream interaction at the apocenter as the major energy dissipation site \citep{Piran2015,Jiang16,Bonnerot17}.  Recently \cite{Dai2018} explain the X-ray/optical dichotomy by considering the inclination effect and the angular distribution of the mass outflow's property such as density, speed and temperature.

Some TDEs have been discovered to have surprisingly low blackbody temperatures of (1 $\sim$ 3) $\times 10^{4}$ K through wide-field UV and optical surveys \citep{Gezari09,Gezari12,van Velzen11,Arcavi2014,holoien14,holoien16a,holoien16b,Blagorodnova17,Hung17}, which can not be explained through radiation from traditional accretion process. These are attributed to larger radiative radii associated with a reprocessing layer \citep{Loeb and Ulmer1997,Guillochon14,Roth16}, which form from a radiatively driven wind \citep{Miller15,Strubbe and Murray15,metzger16} or the radiation from stream-stream collisions during the circularization to form the disk \citep{Lodato12,Piran2015,Shiokawa15,Jiang16,Krolik16,Bonnerot17,Wevers17}. On the other hand, one should expect the X-rays to show up eventually, once the wind subsides or the circularization finishes.

In this paper, we analyse the observations of a recent, nearby TDE candidate AT2019azh which shows a long  (by $\sim$ 200 days) delayed X-ray brightening with respect to its UV/optical peak, a pattern similar to that seen in the past TDE candidate ASASSN-15oi \citep{gezari17}. In \S\ref{sec2} we describe the observations and data reduction in optical, UV and X-ray bands. Then we analyze its multi-wavelength light curves in \S\ref{sec3}, present the spectral energy distribution (SED) fit and show the source's bolometric behavior (evolution of temperature and photospheric radius) in \S\ref{sec4}. We discuss three physical scenarios in TDEs in order to explain its multi-waveband behavior in \S\ref{sec5}. We summarize the results and conclude in \S\ref{sec:con}. 

During the revision of this paper, \cite{Hinkle20} presented their analysis of a larger data set of AT2019azh.

\section{Observations}   \label{sec2} 

\subsection{Optical discovery}\label{optical discovery}

The bright nuclear transient ASASSN-19dj / ZTF17aaazdba (hereafter we will use its IAU name AT2019azh) was discovered by All Sky Automated Survey for SuperNovae \citep[ASSA-SN;][]{shappee14} on UT 2019 Feb 22.02 \citep{brimacombe19} in the center of an E+A galaxy KUG 0810+227 at $z$ = 0.022 (luminosity distance $D= 96$ Mpc). Follow-up spectroscopic observations by NUTS \citep{NUTS19} and ePESSTO \citep{ePESSTO19} show a featureless blue ($T\sim 10^4$ K) spectrum. 

\begin{figure*}
\begin{center}
\includegraphics[width=15.0cm]{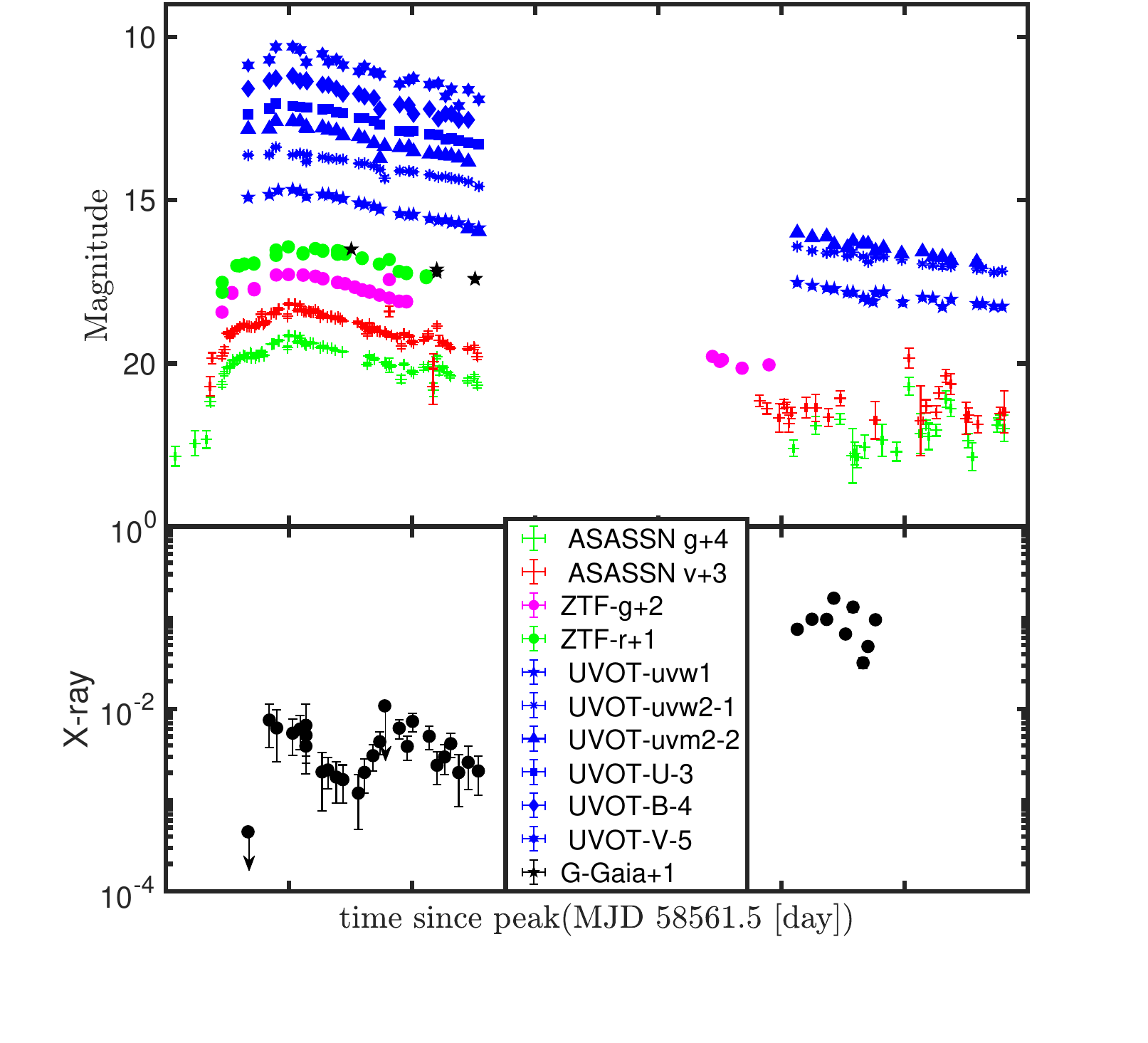}
\caption{Observed light curves of AT2019azh in optical, Swift UV and X-ray bands. The host contribution has already been subtracted in the optical and UV bands ({\it top}). The {\it bottom} panel shows the XRT 0.3-2 keV count rate light curve. The two data points with downward arrows are upper limits. 
}  	\label{fig:mag}
\end{center}
\end{figure*}

\cite{vanvelzen19} reported the Zwicky Transient Facility (ZTF) and \textit{Neil Gehrels Swift Observatory} \citep{gehrels04} UVOT photometry of AT2019azh which show a $\sim$15-day long slowly-rising or plateau phase starting from 2 days after the ASSA-SN trigger. However, later monitoring showed that the flux continued to rise until it peaked at $g$ = 14.4, about 31 days after the trigger (see Figures \ref{fig:mag}). The ZTF and host-subtracted UVOT photometry indicates a temperature of $\log(T)$ = 4.5 $$\pm$$ 0.1, and the ZTF photometry confirms that the transient is consistent with originating from the center of its host galaxy, with a mean offset of 0.07 $$\pm$$ 0.31 arcsec \citep{vanvelzen19}. \textit{Swift} XRT observations on 2019 Mar 11.45 detected 5 soft photons corresponding to a luminosity of $L_X= 2.5\times10^{41}$ erg s$^{-1}$ \citep{vanvelzen19}. 

Based on its nuclear position, persistent blue color, high blackbody temperature and lack of spectroscopic features associated with a supernova or AGN, \cite{vanvelzen19} identified AT2019azh as a TDE, which we will follow hereafter.  

\begin{figure}
\begin{center}
\includegraphics[width=8cm]{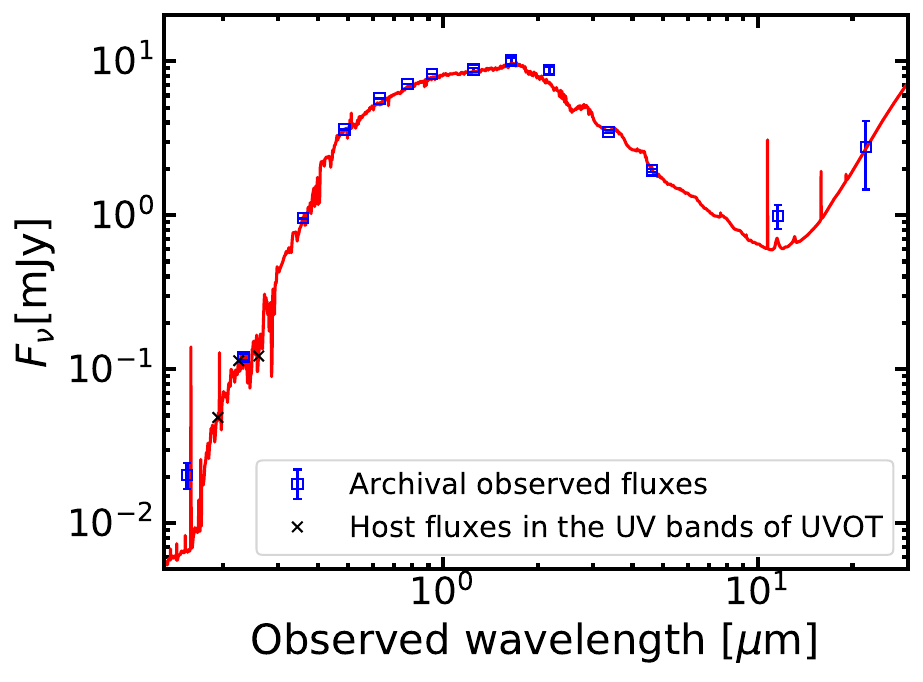}
\caption{ The host galaxy SED fitted by CIGALE with the archival photometry shown in Table \ref{tab:p}, from which we get the host magnitudes in the bands of UVOT-uvw1, uvw2 and uvm2 (black crosses). The blue squares show the archival photometry data.}  	\label{fig:ext}
\end{center}
\end{figure}

We collect the public available ASSA-SN $g$ and $V$ band\footnote{\href{https://asas-sn.osu.edu/light_curves/07988c67-2399-46f1-a9dc-3608c7e8141c}{https://asas-sn.osu.edu/light\_curves/07988c67-2399-46f1-a9dc-3608c7e8141c}}, the ZTF $g$ and $r$ band\footnote{\href{https://lasair.roe.ac.uk/object/ZTF17aaazdba/}{https://lasair.roe.ac.uk/object/ZTF17aaazdba/}}, \textit{Swift} UVOT, \textit{Gaia} $G$ band photometry data. 
For host subtraction in the \textit{ASAS-SN}, \textit{Gaia} and \textit{UVOT-UBV} bands, we take the average fluxes of those quiet periods (listed in Table \ref{tab:q}) before or after the transient as the host fluxes. \textbf{For the UVOT-uvw1, uvw2 and uvm2 band host fluxes, we use CIGALE\citep{Boquien19} to fit the host SED of the galaxy with the archival photometry of \textit{SDSS}, \textit{GALEX}, \textit{MASS} and \textit{WISE}, as is shown in Figure \ref{fig:ext}.} The host photometry that were used are listed in Table \ref{tab:p}. The derived host magnitudes are listed in Table {\ref{tab:host}}. Then the host-subtracted multi-band light curves of the source are plotted in Figure \ref{fig:mag}, and the host-subtracted photometry data are listed in Table \ref{tab:host_subtracted}. \textbf{Based on the  CIGALE fit, KUG 0810+227 has a stellar mass of $M_{S}$ = $1.14^{+0.06}_{-0.06}\times10^{10} M_{\odot}$, an age of $2.83^{+0.005}_{-0.005}$ Gyr and the star formation rate of $2.37^{+0.12}_{-0.12}\times10^{-3} M_{\odot}/yr$}

\textbf{Using the relation between the central BH mass and the total stellar mass of the host(i.e., $M_{BH} = 0.2$$\pm$$0.1\%M_{S}$) in \cite{reines2015}, we estimate the BH mass to be $M_{BH} = 2.3^{+1.3}_{-1.2}\times 10^7 M_{\odot}$, which is consistent with that of \cite{Hinkle20}.}

\begin{table}
\caption{The quiescent periods used for host subtraction and the corresponding magnitudes.}  \label{tab:q}
\begin{tabular}{llll} 
\hline  
Filter    &  & Period (day) & Magnitude \\ 
\hline  
ASAS-SN $g$      & before peak  & 40-55 &  15.20$\pm$0.01\\
ASAS-SN $V$      & before peak  &25-55  &  15.17$\pm$0.01 \\
Gaia            & before peak   &135-1599   &   16.82\\
U            &after peak   &206-421&             16.96$\pm$0.10\\
B            &after peak   &206-416&             15.74$\pm$0.05\\
V            &after peak   &206-421&                15.27$\pm$0.05\\
\hline              
\end{tabular}  
\end{table}
\begin{table}
\caption{Archival \textit{GALEX}, \textit{SDSS}, \textit{MASS} and \textit{WISE} photometry of the host.}  \label{tab:p}
\begin{tabular*}{8cm}{p{4cm}p{1.5cm}p{4cm}}  
\hline  
Filter    &  Magnitude   &   Uncertainty       \\ 
\hline  
$FUV$ (GALEX)               &20.62                        &0.21\\
$NUV$ (GALEX)               &18.70                          &0.05\\
$u$ (SDSS Model)            &16.49                           &0.01\\
$g$ (SDSS Model)            &15.01                            &$-^(1)$\\
$r$ (SDSS Model)            &14.50                            &-\\
$i$ (SDSS Model)            &14.27                            &-\\
$z$ (SDSS Model)            &14.09                           &-\\
$J$ (2MASS/integration)     &13.14                           &0.03\\
$H$ (2MASS/integration).    &12.52                           &0.05\\
$K_s$ (2MASS/integration).  &12.20                           &0.06\\
$W1$ (WISE)                 &12.37                           &0.03\\
$W2$ (WISE)                 &12.36                           &0.03\\
$W3$ (WISE)                 &11.26                           &0.20\\
$W4$ (WISE)                 &8.70                           &0.51\\
\hline  
\end{tabular*}
\begin{tablenotes}
\item[1] (1) means that the uncertainty is less than 0.01.
\end{tablenotes}
\end{table}

\begin{table}.
\caption{The derived host galaxy magnitudes in the \textit{Swift} UVOT bands.}  \label{tab:host}
\begin{tabular*}{5cm}{p{2cm}p{3cm}}
\hline  
Filter    &  Magnituude   \\ 
\hline  
$uvw1$                           &18.69                             \\
$uvw2$                          &19.69                             \\
$uvm2$                         &18.76                             \\
\hline  
\end{tabular*}  
\end{table}

\begin{figure*}
\begin{center}
\includegraphics[width=13cm]{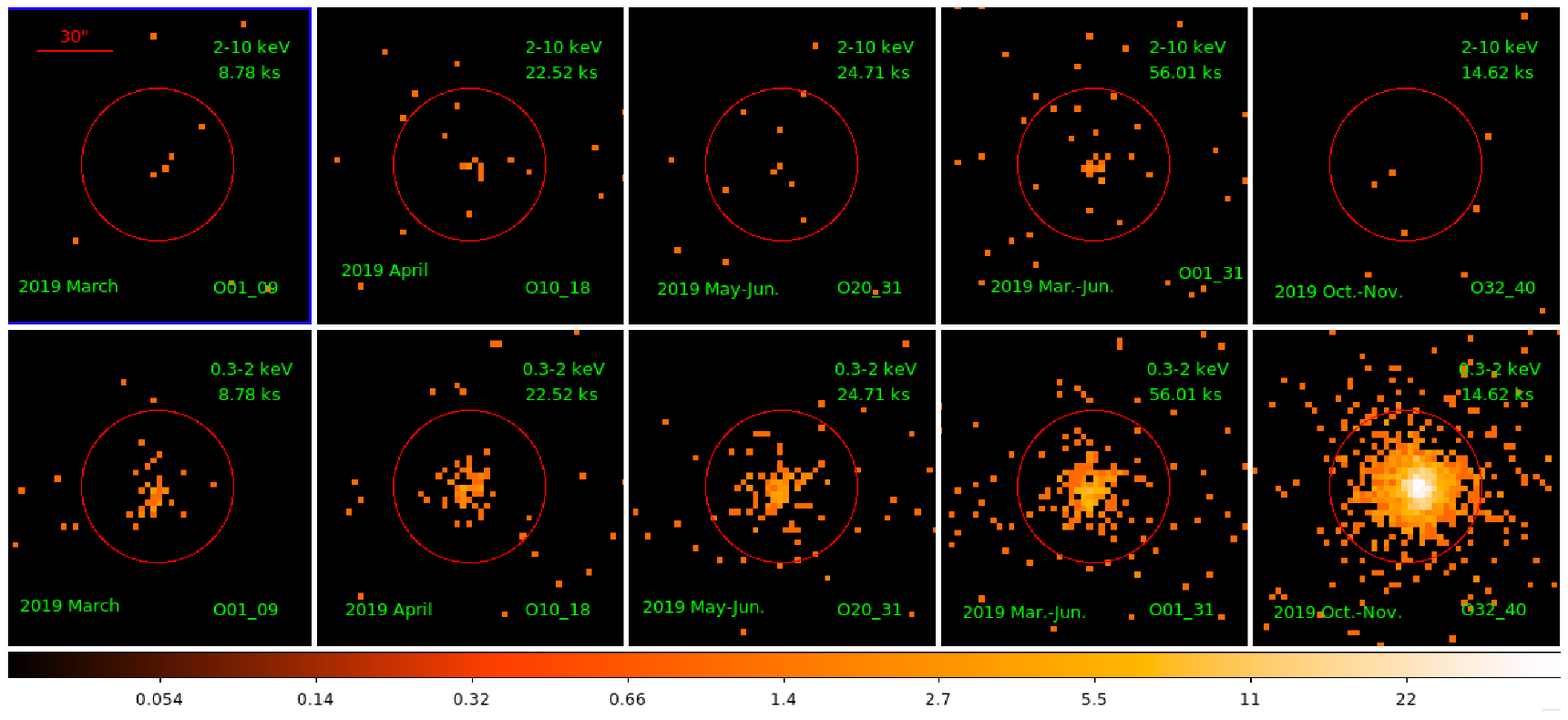}
\caption{Stacking {\it Swift XRT} images in the 2-10 keV (upper panels) and 0.3-2 keV (lower panels) bands.}  
\label{fig:stickingimg}
\end{center}
\end{figure*}
\begin{figure}
\begin{center}
\includegraphics[width=7.5cm, angle=0]{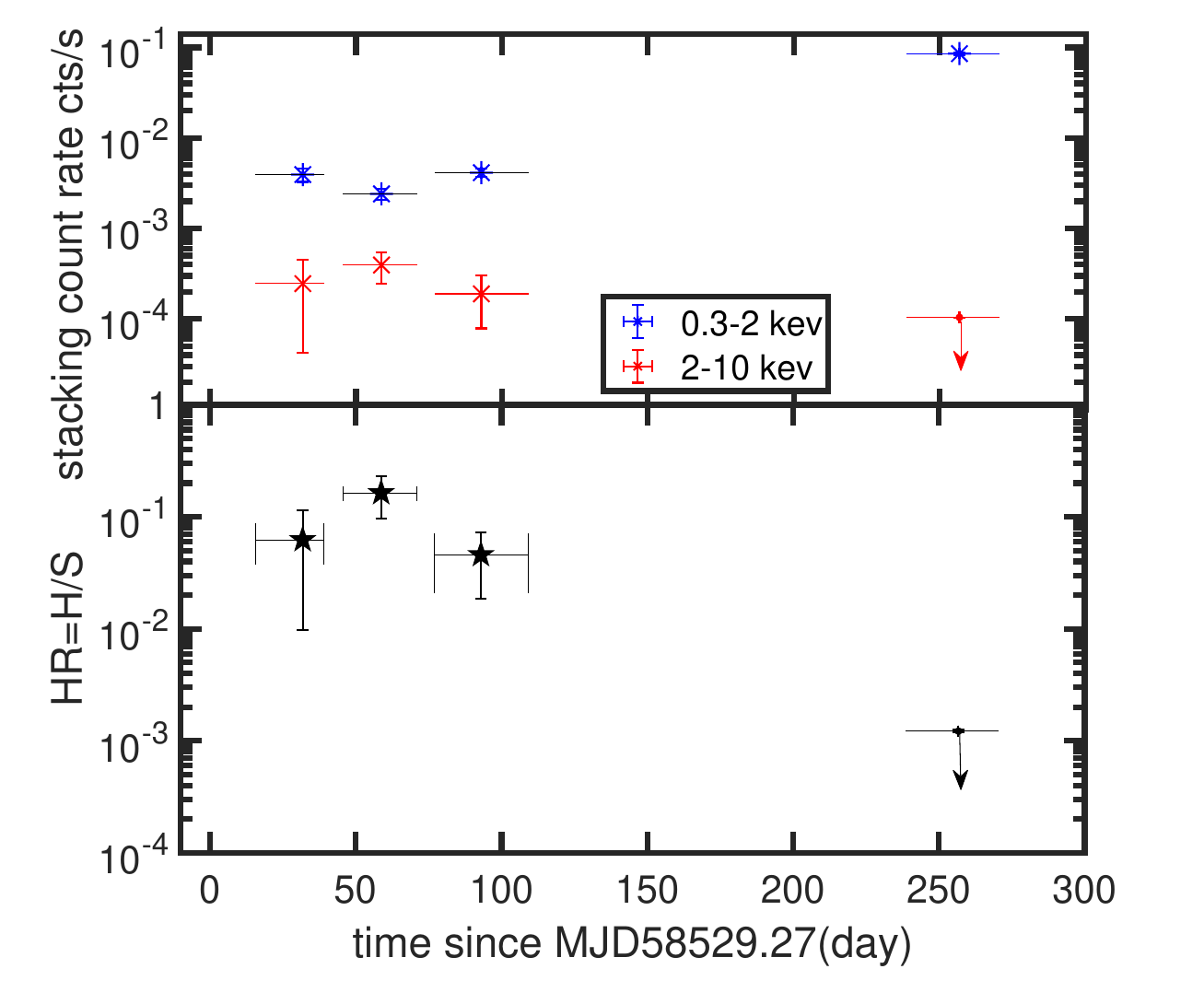}
\caption{The evolution of X-ray hardness of AT2019azh. Asterisks and crosses are the stacking count rates in 0.3-2 keV and 2-10 keV, respectively. The hardness ratio is defined as the count rate ratio between the two bands: $HR$ = (2-10 keV) / (0.3-2 keV).}		\label{fig:x-ray}
\end{center}
\end{figure}

\subsection{Swift UV and X-rays}

{\it Swift} observed AT2019azh for 36 times since March 2, 2019 (update to Nov. 11, 2019). We download and reduce the {\it Swift} XRT \citep{burrows05} data with the software HEASoft V.6.26 and the latest updated calibration files. The UV telescope {UVOT} \citep{roming05} observed the source with multi-wavelength filters (V, B, U, UVM2, UVW1, UVW2). We extract the source photometry from the source region of radius of 3'', and the background from a source-free region with radius of 20'' near the source position , using the task of `uvotsource'. The UVOT photometry results are presented in Figure \ref{fig:mag}. 

For {\it XRT}, we reprocess the event files with the task `xrtpipeline', and select the event files which operated in Photon Counting mode. The source is detected in almost all observations, with a count rate in the range of 0.001 - 0.1 cts s$^{-1}$ in 0.3 - 2 keV. The source file is extracted using a source region of radius of 30''. The background is estimated in an annulus region centered on the source position, with an inner radius of 60'' and outer radius of 90''. Due to the low counts, we only extract the 0.3-2 kV band count rate for single observations. They are listed in Table \ref{tab:1} and plotted in Figure \ref{fig:mag}. 

We also stack the {\it XRT} observations into five groups and list them in Table \ref{tab:1}, which are marked as O01\_09, O10\_18, O20\_31, O01\_31, and O32\_40. The 0.3-2 and 2-10 keV band stacking images for each group are shown in Figure \ref{fig:stickingimg}. We find that the most of X-ray photons are detected in the soft band (0.3-2 keV). Interestingly, about two dozen of X-ray photons are detected in the hard band (2-10 keV) during the observations before June 4, 2019 (O01\_31), and half of them are detected during the 2019 April observations  (O10\_18). We extract the source and background files from the stacking events files. The stacking count rates in 0.3-2 keV and 2-10 keV, respectively, are listed in Table \ref{tab:1}, and are plotted in Figure \ref{fig:x-ray} as well. 

Interestingly, the soft band X-ray count rate shows a factor of $\sim$ 25 increase from O01\_31 to O32\_40, while the hard band X-ray count rate dims to be undetectable, as is shown in Figure \ref{fig:x-ray}. We will come back to this strong hardness evolution feature in \S \ref{sec:X-rays}.
     
\subsection{Radio}		\label{sec:radio}

\begin{figure}[h]
\begin{center}
\includegraphics[width=7.5cm, angle=0]{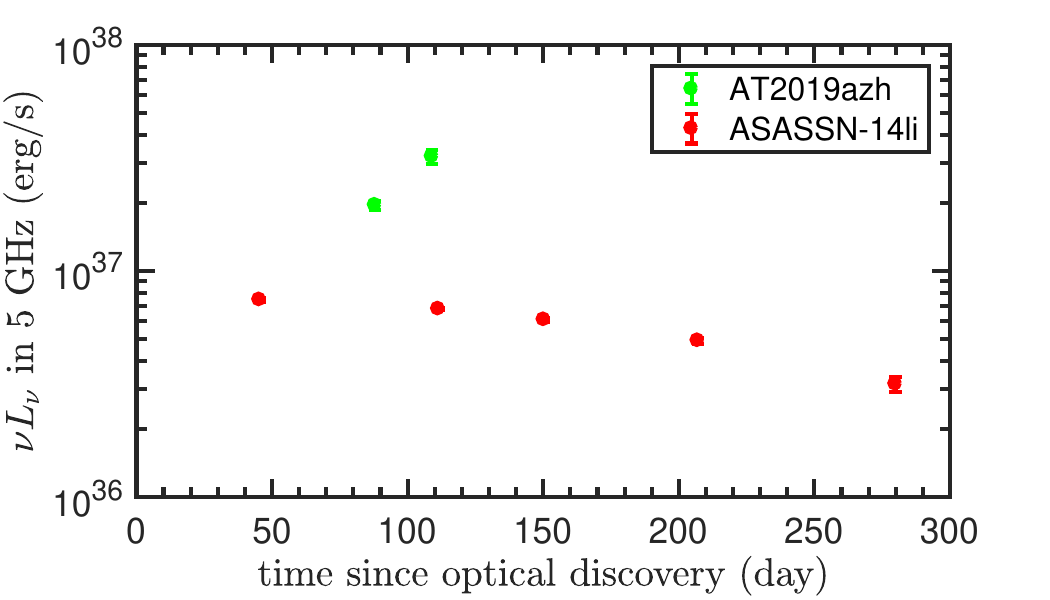}
\caption{Radio light curve of AT2019azh. Data is from \cite{Perez-Torres19}. Its optical discovery date is set to be MJD58536. Another TDE ASASSN-14li \citep[data from][discovery date MJD56983]{Alexander16} is plotted for comparison.}		\label{fig:radio}
\end{center}
\end{figure}

\cite{Perez-Torres19} reported radio detections of the source at 90 - 110 days after the ASAS-SN detection, which we plot in Figure \ref{fig:radio}. For comparison, we also plot the data of another radio bright TDE ASASSN-14li.

The host galaxy of AT2019azh is a non-active galaxy based on the lack of related spectroscopic feature \textbf{\citep{NUTS19,ePESSTO19,vanvelzen19}}. Therefore, its radio emission ($\sim 10^{37}$ erg s$^{-1}$) is unlikely due to AGN activity. It is commonly believed that radio emission from TDEs are produced from the interaction of a relativistic jet or non-relativistic outflow with ambient medium. We discuss this later in \S \ref{sec:cio}.

\section{Temporal Behavior}	\label{sec3}

The ASAS-SN and ZTF data make AT2019azh one of the few TDE candidates with a well-sampled rising light curve.  It rises in the ASAS-SN $g$, $V$ and ZTF $g$, $r$ bands to the peak within 35 days, then decays gradually in all UV / optical bands. In addition, there is a minor re-brightening  in the AS-ASSN $g$ and $V$ bands at $t \approx 50$ d. 

\begin{figure}
\begin{center}
\includegraphics[width=9cm, angle=0]{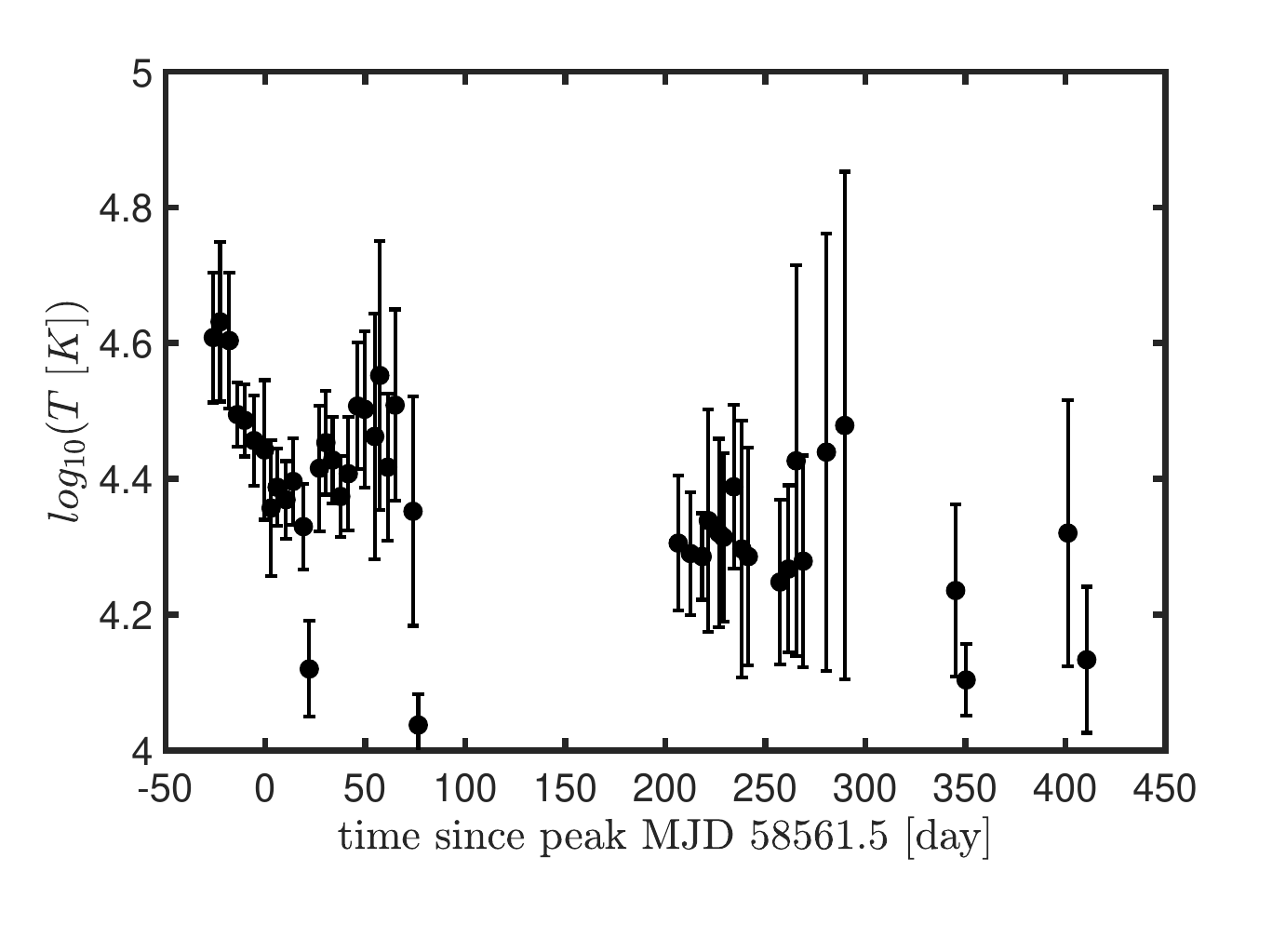}
\caption{The temperature evolution of AT2019azh from blackbody fits to the UV/Optical SED.}\label{fig:T}
\end{center}
\end{figure}

One notable feature in Figure \ref{fig:mag} is that, the data in the several filters indicate very little color evolution in a time span of $\sim$ 100 days, except for the earliest 10 days of the rising. This is supported by the inferred photospheric temperature evolution to be shown later in Figure \ref{fig:T}. It is also in line with the earlier report by \cite{vanvelzen19} and is a characteristic of most previously found TDEs \citep[e.g.,][]{holoien19,vanvelzen20}. 

The X-ray count rate light curve in Figure \ref{fig:mag} shows a very different behavior. Apparently it has two slow flares during the optical bright phase, each varying by a factor of $\sim$ 10 with a duration of $\sim$ 50 days. Notably, the first X-ray detection is $>17$ times brighter than 9 days ago when there was a non-detection. Such an early, sharp variability was also seen in an X-ray TDE AT2018fyk \citep{Wevers19}. Most importantly, the X-ray brightens by a factor of $\sim$ 30 at around 250 days after discovery. This late X-ray brightening is unusual for TDEs, and the only past similar case is ASASSN-15oi \citep{gezari17}.

The behaviors in X-rays and in UV/optical are quite different. The early ($t < 100$ d) X-rays show significant variations while UV/optical are in smooth rise and fall. Later, UV/optical decay monotonically for the rest of the time, but X-rays rise to its peak in about 200 days. These strongly suggest that the two emission components are produced probably at different locations and by different dissipation processes. This is helpful in discerning the most appropriate physical scenario in \S \ref{sec5}.

\begin{figure}
\begin{center}
\includegraphics[height= 6.cm, width=9cm, angle=0]{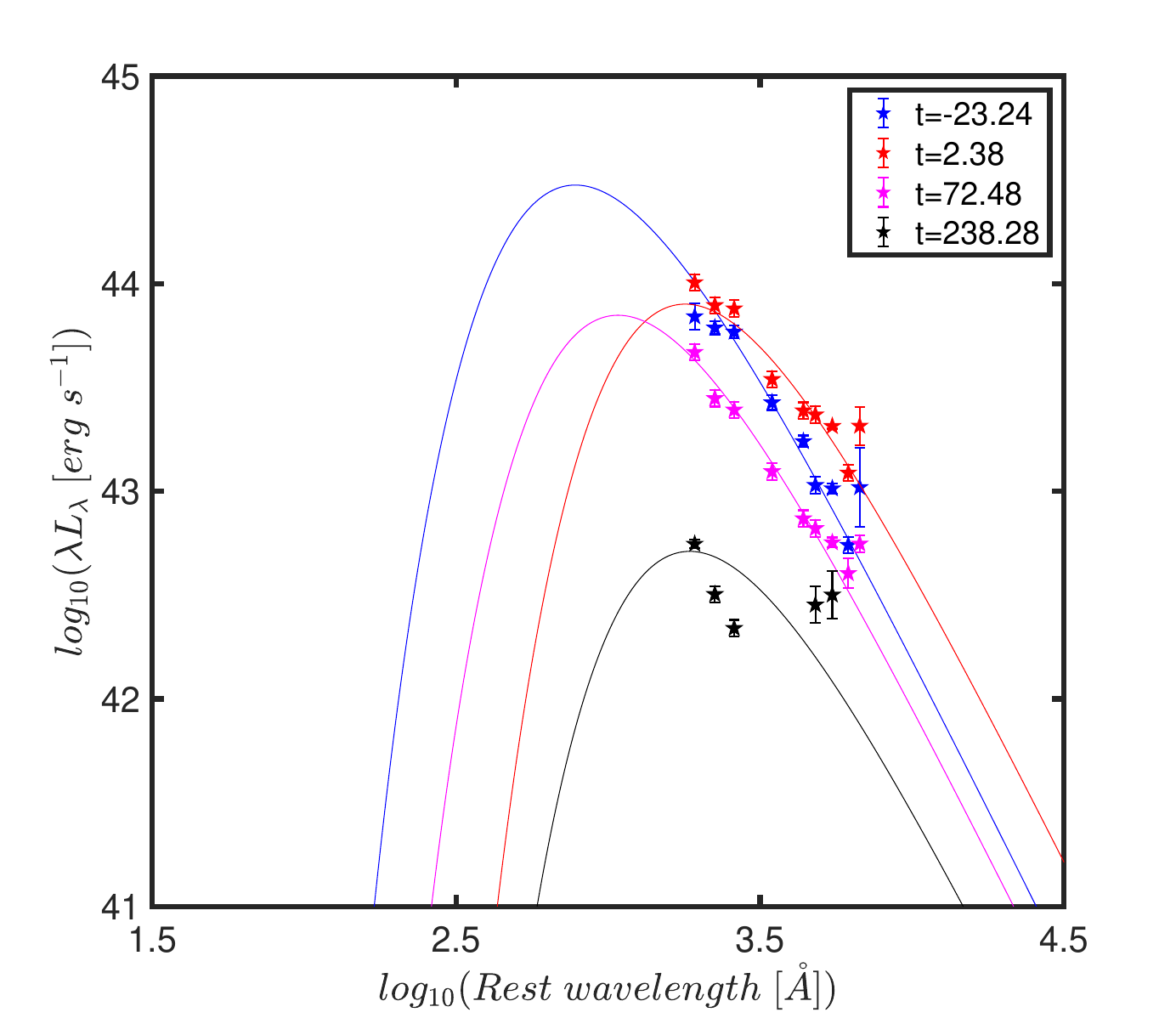}
\caption{Blackbody fits to the multi-epoch SEDs composed of UV and optical photometry data. The numbers mark the times in days since the optical peak. For cleanness, we plot four selected times representing the early, peak, decaying and late epochs, respectively.}     	\label{fig:sed}
\end{center}
\end{figure}

\begin{figure}
\begin{center}
\includegraphics[width=9cm, angle=0]{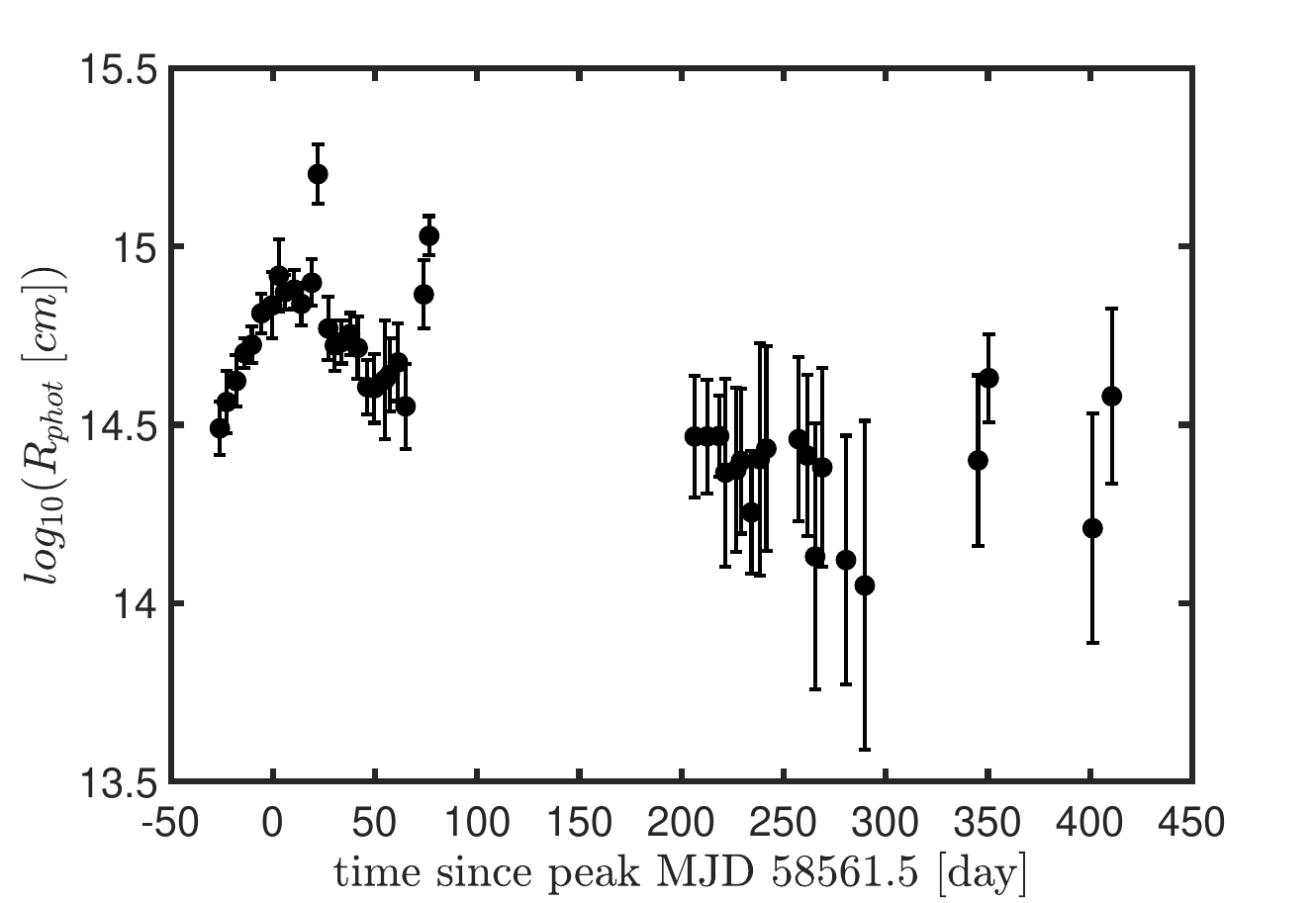}
\caption{Evolution of the photospheric radius derived from the blackbody SED fits to the UV/Optical. }
\label{fig:R}
\end{center}
\end{figure}

\section{Spectral analysis}		\label{sec4}

\subsection{UV / optical}

We assume an extinction law same as that in \cite{Cardelli1988} with $R_V = 3.1$ and Galactic extinction\footnote{\href{https://irsa.ipac.caltech.edu/applications/DUST/}{https://irsa.ipac.caltech.edu/applications/DUST/}} of $A_V= 0.118$ \citep{Schlafly2011}. \textbf{Additionally, we assume there is no host extinction.} We fit the SED with the blackbody model by the SUPERBOL \citep{nicholl2018} which uses least square method. We adopt 30000 K as the prior of temperature.

\begin{figure*}
\begin{center}
\includegraphics[width=18cm, angle=0]{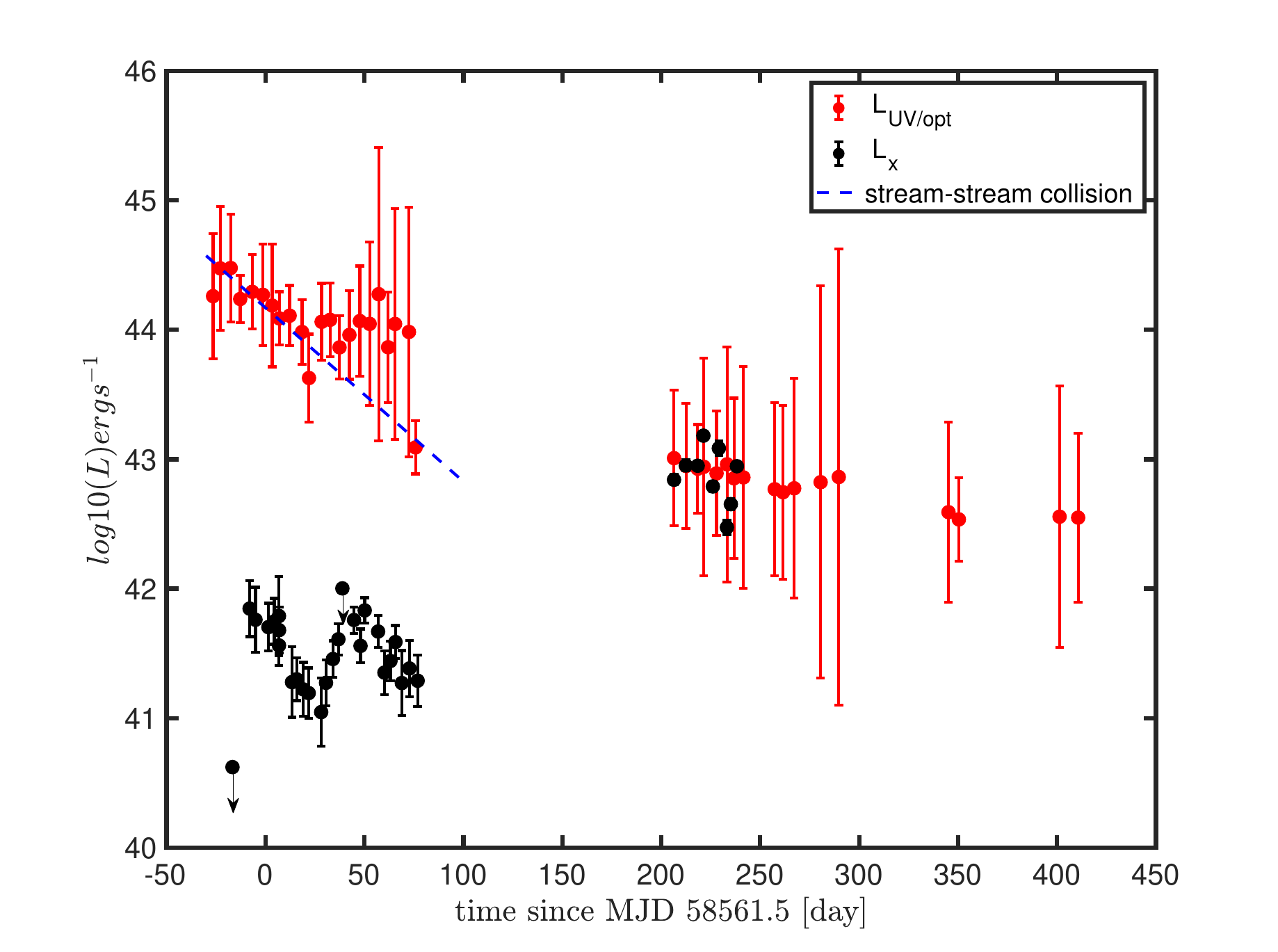}
\caption{UV/optical luminosity evolution of AT2019azh from the blackbody SED fits (red) and X-ray luminosity evolution from Swift (black). The blue dotted line shows the fitting result of Eq. \ref{eq:l}.}		\label{fig:L}
\end{center}
\end{figure*}

The temperature $T$ shown in Figure \ref{fig:T} generally decreases fast near the light curve peak and rises at about 20 days after peak. Then at later ($t >$ 200 d) times $T$ roughly shows a slow drop. This later trend is roughly consistent with \cite{Hinkle20} but is inconsistent with \cite{vanvelzen20} who find $T$ rises at later times. Additionally, the evolution of the photospheric radius for the UV/optical is shown in Figure \ref{fig:R}, which has the same evolution as \cite{Hinkle20}, however our result shows a bit larger radius than that of \cite{Hinkle20}. The bolometirc luminosity, shown in Figure \ref{fig:L}, rises to peak at $\sim 10^{44.5}\;erg s^{-1}$, which is consistent with \cite{vanvelzen20} but a bit lower than \cite{Hinkle20} (i.e.,$\sim 10^{45}\;erg s^{-1} $).


\begin{figure*}
\centerline{
\includegraphics[width=0.4\textwidth]{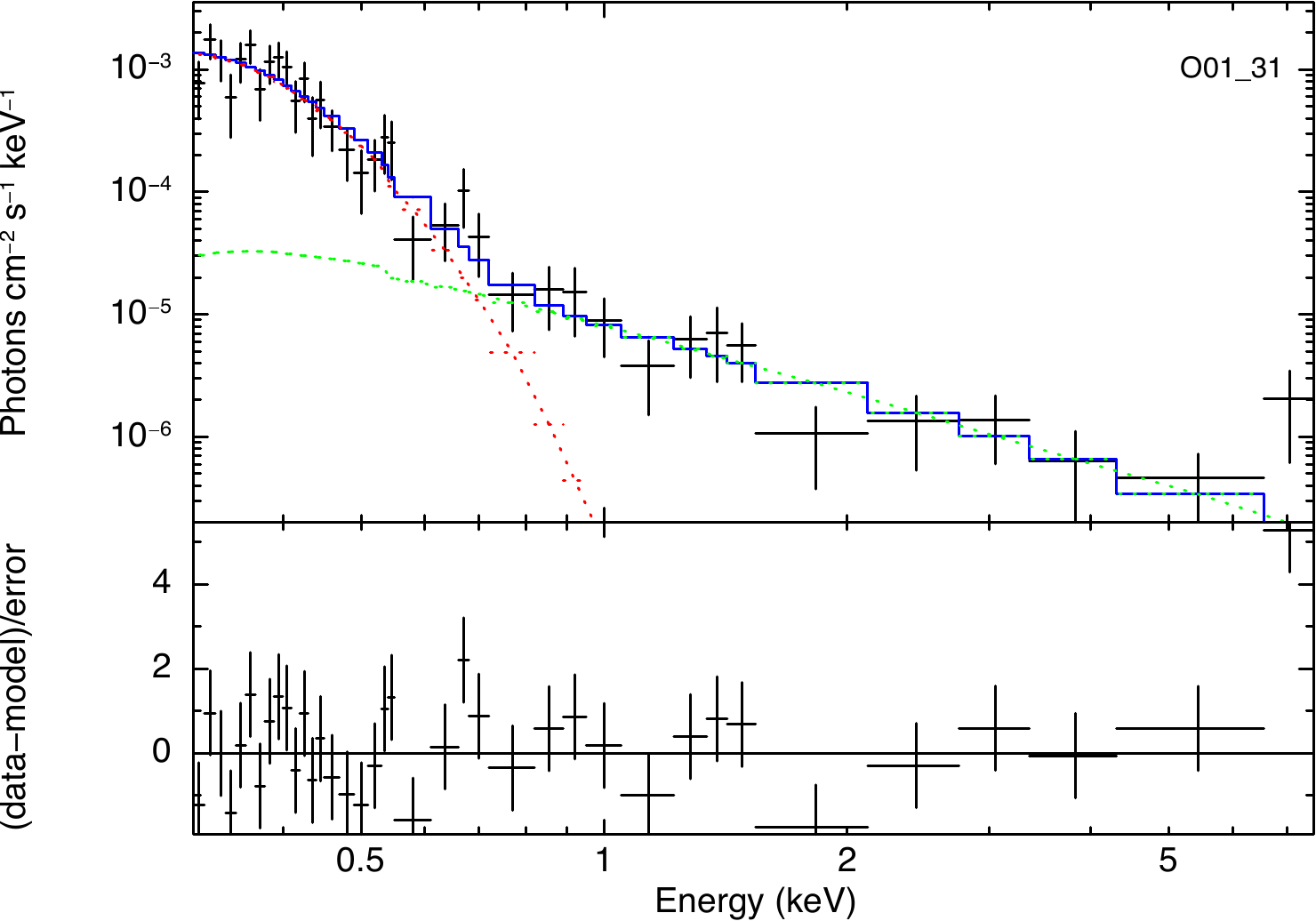}
\includegraphics[width=0.4\textwidth]{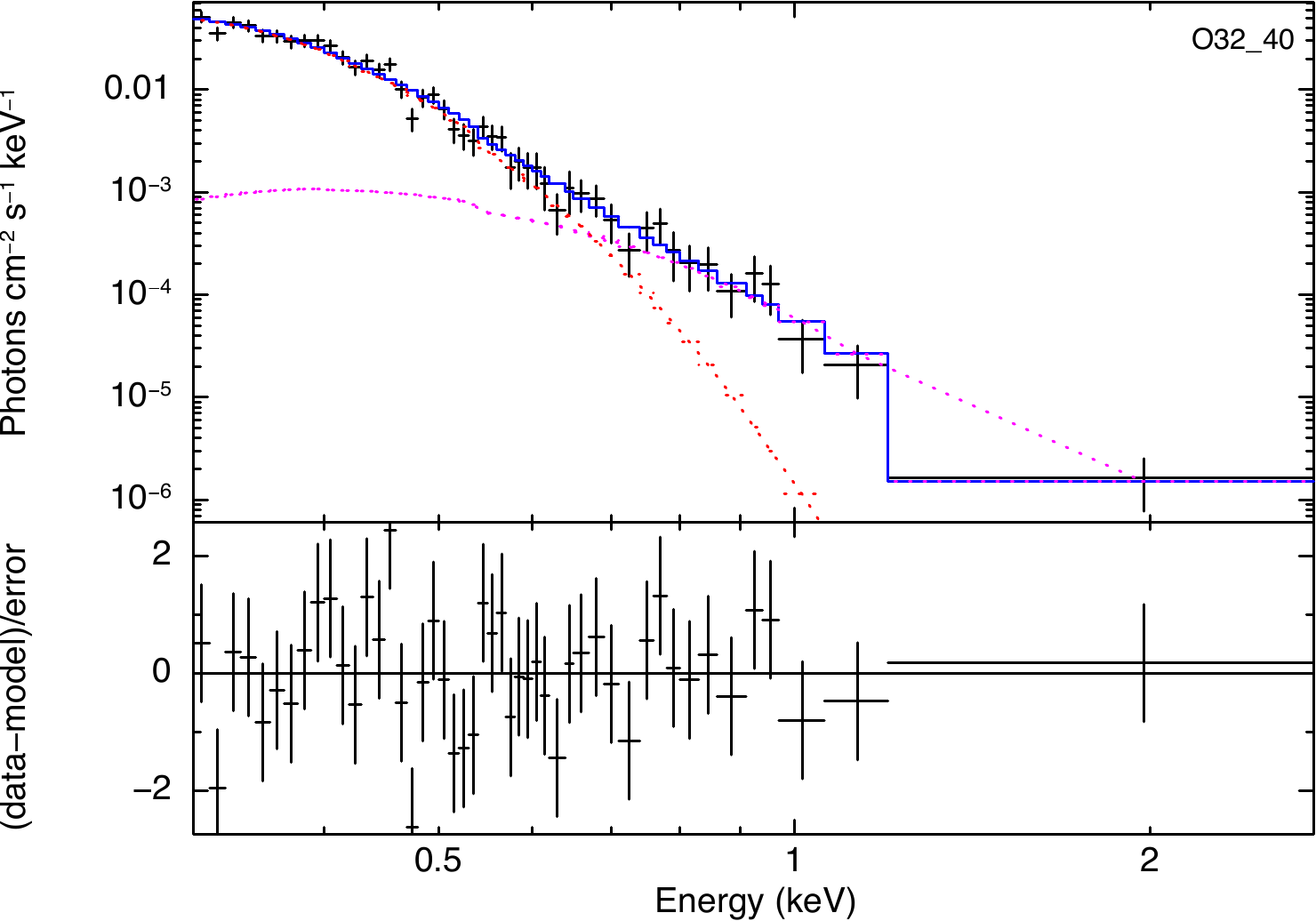}
}
\caption{Stacking {\it Swift XRT} spectra and best fit models. The data is marked as black color, the best fit models are marked as blue color, the blackbody model component is marked as red dot line (the second blackbody in O32\_40 is marked as pink dot line) and the power-law model component is marked as green dot line.}    \label{fig:xrayspec}
\end{figure*}

\subsection{X-rays}	\label{sec:X-rays}

For Swift X-ray data, we fit the two stacking spectra which are grouped in before and after June 4, 2019 (O01\_31 and O32\_40). We group the data to have at least 4 counts in each bin, and adopt mainly the C-statistic for the Swift spectral fittings which are performed using XSPEC (v.12.9; Arnaud 1996). We try to fit the spectra with five different models, which are a single power-law, a double power-laws, a single blackbody, a double blackbodies, and a single power-law plus single blackbody, respectively. For Galactic absorption, we adopt a column density of $N_{H}$ = 4.15 $\times$ $10^{20}$ cm$^{-2}$ \citep{HI4PI Collaboration2016} in the direction of AT2019azh.

The fitting results are listed in the Table \ref{tab:2}. We find that the best fit and accepted model for O01\_31 is a single power-law, with photon index $\Gamma = 1.9 $$\pm$$ 0.6$, plus a single blackbody of temperature $kT=56 $$\pm$ 9 eV.
Both double power laws and a power law plus a black-body can be accepted by the spectral fitting for the spectrum of O01\_31. However, the photon index of steeper PL component in double PL model ($\Gamma \sim 6.1$) is too steep, even steeper than some cases of AGN with extremely soft X-ray emissions (e.g., RX J1301.9+2747, \cite{sun13}). Therefore, we prefer the later model. Since the dominated component is the black body rather than the PL, we have change the model name to a `black body plus PL' model. 

Based on the best fit results, we estimate the X-ray fluxes of the two stacking spectra. The 0.3-2, and 2-10 keV band unabsorbed fluxes in O01\_31  are $2.34^{-0.37}_{+0.10} \times 10^{-13}$ erg cm$^{-2}$ s$^{-1}$,  and $2.5^{-0.8}_{+1.0} \times 10^{-14}$ erg cm$^{-2}$ s$^{-1}$, respectively.The 0.3-2 keV band unabsorbed flux in O32\_40  is $7.10^{-0.50}_{+0.08} \times 10^{-12}$ erg cm$^{-2}$ s$^{-1}$. Note that there are two blackbody components in O32\_40, however, the 0.3-2 keV band flux of the component with higher temperature is only $\sim 7\%$ of the flux of the one with the low temperature in O32\_40. Using the stacking spectral fitting results, we convert the count rate of each XRT exposure to the 0.3 - 10 keV flux. Replacing the blackbody component in the best fit models with a multi-blackbody from accretion disk model ('diskbb'), we also derive the  X-ray blackbody radii, which are $\sim 7 \times 10^{10}$, $6\times 10^{11}$ and $1\times 10^{10}$ cm, respectively, for the blackbody component in O01\_31, the first, and the second one in O32\_40. These radii are 0.07-4 times of $R_g$ ($R_g = GM /c^2$), if considering an SMBH mass of $10^{6} M_{\sun}$.

As was shown in Figure \ref{fig:x-ray},  the X-ray hardness ratio drops substantially toward later times and becomes softer when the source is brighter, which is reminiscent of the state transitions behavior typically seen in BH X-ray binaries \citep[e.g.,][]{remillard06}. Thus, the X-rays here can be considered as the signature of the BH accretion. Therefore, the early X-rays probably correspond to a low accretion state which is harder, whereas the late X-ray brightening at $t=$210 d corresponds to a high accretion rate where the hard photons disappear. 

\subsection{Comparison between UV/optical and X-rays}

\begin{figure}
\begin{center}
\includegraphics[width=9cm, angle=0]{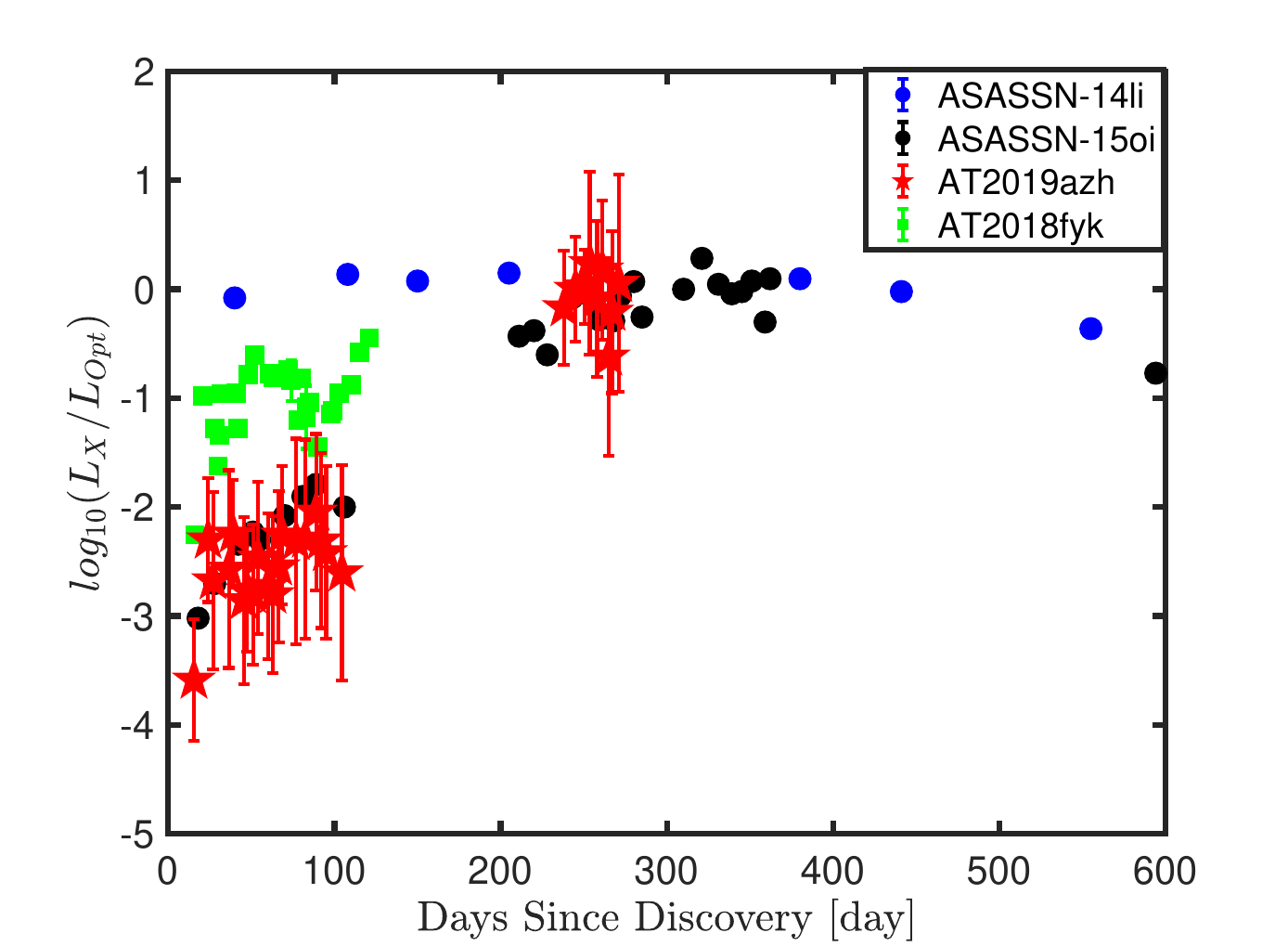}
\caption{Evolution of the luminosity ratio of X-rays over UV/Optical. The data source for the other three TDEs: ASASSN-14li and ASASSN-15oi are from \cite{gezari17} and AT2018fyk is from \cite{Wevers19}.}\label{fig:LxL}
\end{center}
\end{figure}

Figure \ref{fig:L} shows the UV/optical and X-ray luminosity evolution of AT2019azh. In the early time($t \leq 100\;day$) the X-ray has two peaks at 0 and 50 day, respectively, while UV/optical decreases, suggesting the lack of a correlation between the X-ray and UV/optical. In addition, we plot the evolution of the X-ray-to-UV/optical luminosity ratio $L_X/L_{\rm opt}$ in Figure \ref{fig:LxL}. It shows that AT2019azh and ASASSN-15oi are generally very similar, in that this ratio rises from 0.001-0.01 during the early optically bright phase to $\sim 1$ later. They differ from  ASASSN-14li which shows almost constant $L_X/L_{\rm opt} \sim 1$.


\section{Physical interpretation}			\label{sec5}


The characteristic timescale for a TDE is set by the orbital period of the most tightly bound debris, known as the fallback time \citep{chen and shen21}
\beq\label{eq:tfb}
t_{\rm fb} = 41\beta^{-3}M_6^{1/2}r_*^{3/2}m_*^{-1} \; d
\eeq 
where $M = M_6 \times10^{6}$ $M_{\odot}$ is the mass of the black hole, $R_*=r_{*}\times R_{\odot}$ and $M_{*} = m_{*}\times M_{\odot}$ is the star's radius and mass, $\beta = R_{t}/R_{p}$,\;$R_{t} = R_{*}(M/M_{*})^{1/3}$ is the star's tidal disruption radius and $R_p$ is the star's initial orbital pericenter radius \citep{Rees1988,Phinney1989}. After disruption, the less bound debris follows the most bound debris in returning, in a rate that drops with time as \citep{Rees1988,Phinney1989,Lodato09,RamirezRuizRosswog2009,Guillochon13}
\beq		\label{eq:mdot}
\dot{M}_{fb}\simeq\dot{M}_{peak} \left({\frac{t}{t_{\rm fb}}}\right)^{-5/3}.
\eeq
In the following we discuss three possible scenarios for interpreting the UV/optical and X-ray behavior of AT2019azh.
 
\subsection{Super-Eddington accretion for the UV/optical peak}	 	\label{sec:super}
  
The optical bright stage could be super-Eddington which peaks in the UV/optical bands, during which various energy dissipation processes will produce winds or outflows \citep{StrubbeQuataert2009,LodatoRossi2011,metzger16} which will regulate the luminosity \citep{KrolikPiran2012}, and block the X-rays or reprocess them into UV/opticals \citep[e.g.,][]{Dai2018}. After the accretion rate drops below the Eddington rate, the bolometric luminosity falls and X-rays from the inner accretion disk start to be seen \citep{Chen and Shen18}. This transition time can be estimated from Eq. (\ref{eq:mdot}) as 
\beq\label{eq:tedd}
t_{\rm Edd} \simeq 2.2 \,\eta^{3/5}_{0.1}M^{2/5}_{6}r^{3/5}_{*}m^{1/5}_{*}~\mbox{yr},
\eeq
where $\eta= 0.1 \times \eta_{0.1}$ is the efficiency of converting accretion power to luminosity. AT2019azh is visible in X-rays during the early, optical bright phase. This suggests that, in \cite{Dai2018}'s model, the line of sight is somewhat close to the pole direction. 


However, there are two evidences against this early super-Eddington accretion scenario. First, the X-rays and UV/optical light curves of AT2019azh behave very differently during the early time ($t <$ 100 d), as was shown in Figure \ref{fig:mag} and mentioned in \S \ref{sec3}. They show no sign of temporal correlation between the two bands that one should expect to see in this scenario, since under which at the super-Eddington accretion stage, the two bands are produced by the same accretion process. 

At this stage, both X-rays and UV/optical are emitted from the accretion-induced outflow, but the two wavebands are from different latitudes whose out-flowing matter densities are quite different, so are the depths of their photospheres (or the last scattering surfaces), which causes the difference of the photon energies of the two components. Under this interpretation, the short-term variability, like the two flares seen in the early X-rays of AT 2019azh, are most likely caused by some variation in the central accretion rate, which would modulate the out-flowing mass rate as well. Thus, a temporal correlation between X-ray and UV/opt is expected.\footnote{We notice that the early UV/optical and X-rays in AT2018fyk do show a weak temporal correlation \citep{Wevers19}. Therefore, AT2018fyk may belong to the early super-Eddington accretion scenario.}.

Second, as was shown in Figure \ref{fig:x-ray} and mentioned in \S \ref{sec:X-rays}, hard X-ray photons appeared during early time but disappeared later. The resemblance of this particular behavior to the state transition pattern of BH X-ray binaries suggests that those X-rays, both the early and the late, are signatures of accretion. The early X-rays probably correspond to a low accretion state which is harder, whereas the late X-ray brightening at $t \simeq$ 200 d corresponds to a high accretion rate so the hard photons disappear. This is in sharp contradiction with the early UV/optical peak being a super Eddington accretion phase. 


\subsection{Stream-stream collision followed by delayed accretion}	\label{sec:acc}

Next we consider a two-process scenario, in which the late X-ray brightening comes from the delayed accretion through a recently formed accretion disk, and the early UV/optical emission is from the stream-stream collisions \citep[e.g.,][]{Piran2015,Jiang16} before the major body of the accretion disk is formed.

For generality, we consider both full ans partial disruption cases. Following \cite{chen and shen21}, after the disruption, the most bound material has a specific energy
\beq		\label{eq:eps0}
\epsilon_0= \frac{GM}{R_t^2} R_* \times \begin{cases}
1, & \text{for} ~\beta \ge \beta_d ~\text{(full TDE)}, \\
\beta^2, & \text{for} ~\beta < \beta_d ~\text{(partial TDE)}.
\end{cases}
\eeq
and an orbital period $t_{fb}$, where $\beta_d$ is the critical penetration factor that separates partial disruptions from full ones. Its orbital pericener radius is equal to that of the star, semi-major axis is $a_0 = GM / (2\epsilon_0)$, eccentricity is $e_0 = 1 - R_p/a_0$, and specific angular momentum squared is $j_0^2= GM R_p(1+e_0)$.


Consider only the "main stream", which is made of the most bound materials, Due to apsidal precession, this stream crosses and collides with itself in successive orbits, reducing its energy. \cite{Bonnerot17} has analytically studied the self-crossing process. Following them and \cite{chen and shen21}, we can write
\beq		\label{eq:diff}
\dot{\epsilon}  = \frac{\dot{\epsilon}_0}{e_0^2} \left(1 - \frac{\epsilon}{\epsilon_c}\right) \left(\frac{\epsilon}{\epsilon_0}\right)^{3/2},
\eeq
where $\dot{\epsilon}_0= \Delta \epsilon_0/t_{\rm min}$ is the dissipation rate in the first orbit, $\epsilon_c \equiv \epsilon_0/(1-e_0^2)= GM / (2R_c)$ is the final value of specific energy and $R_c= R_p (1+e_0)$ is the so-called circularization radius. As the stream progressively dissipates its energy through crossing, $\epsilon$ increases from the initial value $\epsilon_0$ to the end value $\epsilon_c$. Early on, the factor $(1-\epsilon/\epsilon_c) \approx 1$ [\textbf{$\epsilon_0/\epsilon_c = 1 - e_0^2 = 1 - (1 - 2\beta(M_*/M)^{1/3})^2$, for AT2019azh this ratio is approximately equal to 0.018}] varies slowly, so $\dot{\epsilon}$ increases with time. Later when $\epsilon$ approaches $\epsilon_c$, this factor drops quickly, so does $\dot{\epsilon}$. 

Eq. (\ref{eq:diff}) can be numerically solved, but one can get an insight to its behaviour by simplifying it into asymptotic forms according to the early and late regimes, respectively:
\beq
\dot{\epsilon}  \simeq  \dot{\epsilon}_0 \times 
\begin{cases}
(\epsilon/\epsilon_0)^{3/2}, & \text{for}~ \epsilon \ll \epsilon_c, \\
(\epsilon_c/\epsilon_0)^{3/2}  (1 - \epsilon / \epsilon_c),  & \text{for}~ \epsilon \approx \epsilon_c.
\end{cases}
\eeq
Solving these we get the following temporal forms
\beq		\label{eq:dedt-solution}
\dot{\epsilon}(t) 
\begin{cases}
\simeq  \dot{\epsilon}_0  (1-t/t_{\rm cir})^{-3}, & \text{for}~ \epsilon \ll \epsilon_c, \\
\propto \exp[-2 (\epsilon_c / \epsilon_0)^{1/2}(t /t_{\rm cir})],  & \text{for}~ \epsilon \approx \epsilon_c,
\end{cases}
\eeq
where 
\begin{align}\label{eq:tcir}
t_{\rm cir}&= 8\beta^{-1}M_6^{-5/3}m_*^{-1/3}r^2_*t_{fb}\\
&=328\beta^{-4}M_6^{-7/6}r_*^{7/2}m_*^{-4/3}\;d\\
\end{align}
\textbf{For AT2019azh, taking the estimated values below, we obtain $t_{cir} = \;480\;d$, $\epsilon_0 = 1.8\times10^{17}\;erg/g$ and $\epsilon_c = 1.0\times10^{19}\;erg/g$}.

Assuming an efficient radiative cooling process, the luminosity is equal to the dissipation rate times the mass $\Delta M$ of the "main stream"
\beq\label{eq:l}
L = \Delta M\dot{\epsilon}(t) 
\eeq

Letting $\Delta M$ be a free parameter, we fit the early luminosity data (i.e., $t= -20 \sim 20$ d) in Figure \ref{fig:L} with Eq. \ref{eq:l} (only for $\epsilon \simeq \epsilon_c$, i.e., the exponentially decay part), as is shown by the blue dotted line in Figure \ref{fig:L}. Taking $\beta = 0.6$ as an exemplar value (i.e., we suppose that AT2019azh is a partial disruption event which will be discussed later), adopting the mass-radius relation for high-mass main sequence stars $r_* = m_*^{0.6}$ (for $1\textless m_* \textless 10$) \citep{KippenhahnWeigert1994}, \textbf{the fitting gives $m_*\simeq4.6^{+1.4}_{-0.6}$}. The early luminosity decay of AT2019azh is somewhat shallow, as shown in Figure \ref{fig:L}, which suggests a large $t_{\rm cir}$ according to the second line of Eq. (\ref{eq:dedt-solution}). Then from Eq. (\ref{eq:tcir}) it means the fit tends to give small $\beta$ and large $m_*$. 

\textbf{It is unexpected to obtain such a large $m_*$, which is a small probability event compared to the $m_*$ which is less than $1$. However, a smaller $\beta$ means a greater probability of occurrence, which can neutralize the low probability corresponding to $m_*$ to some extent.}

This stellar mass is much higher than the value ($m_*=0.1$) that \cite{Hinkle20} estimated from MOSFiT. However, one should notice that MOSFiT does not consider the process of stream-stream collision and it is built on the hypothesis that UV/Optical emission comes directly from accretion, in contrast with the more realistic scenario that we consider here, i.e., the early UV/Optical peak of AT2019azh is produced by the stream-stream collision.


Following \cite{chen and shen21}, the peak luminosity which is producing by stream-stream collision is
\beq
L_p \simeq 3\times 10^{44} \left(\frac{\Delta M}{M_*/2}\right) m_*^{5/2} r_*^{-9/2} M_6^2 \beta^{9/2}~ \text{erg~s}^{-1}.
\eeq
It scales strongly with the stellar average density, black hole mass and the plunging depth, such that a denser star, a heavier black hole or a deeper plunge will have more luminous signal from streaming crossings, and it may exceed Eddington luminosity for $M > 10^6~M_{\odot}$.
From Figure \ref{fig:L} we obtain  $L_p=10^{44.5}$ as well as the estimated value of $M_{BH}$ and $m_*$, we have:
\beq
\frac{\Delta M}{M_*/2} = 0.0878
\eeq
is roughly consistent with the simulation result $\Delta M = 0.0508M_*/2$ \citep{Guillochon13} and indicates the star was partially disrupted in the encounter. This conclusion is different from \cite{Hinkle20}, who assume complete disruption under the accretion scenario that MOSFiT is based upon.


What produced the early ($t <$ 100 d) low-level X-ray activity? During the stream-stream collisions, the hydrodynamic numerical simulation by \cite{Shiokawa15} shows that some minor amount of debris lost a significant fraction of their angular momentum, such that their pericenter radius could shrink significantly and they could form an early, low mass accretion disk. If this is possible, then one naturally expects that the early accretion rate is low (i.e., sub-Eddington). The hard X-ray photons appearing during this phase might suggest a hot corona is formed during this `low hard' state, similar to what happens in BH X-ray binaries.

The above scenario is naturally consistent with the absence of temporal correlation between early X-rays and UV/opticals, since they are produced by different dissipation processes and at very different locations. 

Once the stream-stream collision process was over, the major body of the disk should have formed, at around $t\sim$ 200 d. The accretion rate has risen to the peak, so does the soft X-ray flux, while hard X-ray photons disappear, probably with the hot corona.


\subsection{CIO's reprocessing of X-rays}   \label{sec:cio}

\cite{Lu and Bonnerot} argue that during the stream-stream collisions, a considerable number of the shocked gas will become unbound and ejected as the so-called collision-induced outflow (CIO), which could reprocess early X-rays (presumably coming from an inner disk which is formed in the same way as was described in \S \ref{sec:acc}) into optical bands. 

This scenario is disfavored for the UV/optical peak in AT2019azh due to the following reasons. If the content of CIO is large and massive so that the CIO could cover the whole sphere around the source, then the early X-rays should not be visible, which is clearly not the case. If the CIO coverage is partial and it does not fully block the line of sight, then one should expect: 1) the observed X-ray flux  shall exceed or at least be comparable to the UV/optical flux, not the opposite, because only a portion of $L_X$ gets reprocessed and becomes $L_{\rm opt}$; 2) the UV/optical should show a temporal correlation with the early X-rays, which is in contradiction with what the early light curves show in Figures \ref{fig:mag} and \ref{fig:L}.

Note that AT2019azh is detected in radio at $t \sim$ 100 d (see \S \ref{sec:radio} and Figure \ref{fig:radio}). This suggests that CIO may actually exist and its interaction with ambient medium produced the radio emission \citep{Lu and Bonnerot}. However, the CIO's reprocessing of X-rays can not be the origin of the UV/optical peak.  
 
\section{Summary and Conclusion}		\label{sec:con}

We present and analyze a large data set by ASAS-SN, ZTF, {\it Swift} and Gaia of the light curves of TDE candidate AT2019azh in optical/UV and X-ray bands. We highlight a rare case in which the late X-rays brightened by a factor of $\sim$ 30-100 around 200 days after discovery, while the UV/opticals continuously decayed. The early X-rays show two flaring episodes of variation, temporally uncorrelated with the early UV/opticals. In addition, we present the evolution of temperature and photospheric radius from the fitting of SED. We found a clear sign of evolution of the X-ray hardness ratio which drops substantially toward later times and becomes softer when the source is brighter.

The drastically different temporal behaviors in X-rays and UV/opticals suggest that the two bands are physically distinct emission components, and probably arise from different locations. The hard X-ray (2 -10 keV) photons found during $t <$ 100 d suggest that the early X-rays must be of accretion origin as well. 

Putting all pieces together, we conclude that the full data are best explained by a two-process scenario, in which the UV/Optical peak is produced by the stream-stream collisions during the circularization phase; the same process causes some low angular momentum, shocked gas to form an early, low-mass accretion disk which emits the early X-rays. The major body of the disk is formed after the circularization finishes, at $t\sim$ 200 d. Its enhanced rate of accretion toward the black hole produces the late X-ray brightening.       

AT2019azh is the second case, after ASASSN-15oi \citep{gezari17}, of TDEs that shows a clear sign of delayed accretion. However, the early detection and full multi-waveband coverage make AT2019azh the first strong case that the emission signature of stream-stream collision is identified and early steps of disk formation can be inferred. At the time of the paper is written, AT2019azh is still detectable in X-rays, so deeper and broader understanding of this event is reachable.  

\acknowledgments
The authors are very grateful to the anonymous referee for the valuable comments and constructive suggestions that helped to improve the quality of the manuscript. This work is supported by the National Natural Science Foundation of China (12073091), Guangdong Basic and Applied Basic Research Foundation (2019A1515011119) and Guangdong Major Project of Basic and Applied Basic Research (2019B030302001).
  


\begin{table}
\caption{The log of {\it Swift}-XRT observation.}  \label{tab:1}
\begin{tabular*}{18cm}{p{1.5cm}p{3cm}p{1.5cm}p{2cm}p{2.5cm}p{2.5cm}p{2.5cm}}  
\hline  
ObsID    &  ObsDate   &      Exposure       &      Count rate      &     Flux \\ 
(11186-)&&&0.3-2 keV&0.3-2 keV\\
   &                 &          (ks)                &   (10$^{-3}$ cts/s)   & ($10^{-13}$erg/s/cm$^2$)  \\ 
\hline  
001  &   2019-03-02T18:00:36 & 2.198         &        $\textless$0.45   &$\textless$0.4\\
003  &   2019-03-11T10:42:36 & 0.529         &        7.55$\pm$3.78     &6.2$\pm$3.16\\
004  &   2019-03-14T02:26:34 & 0.484         &        6.19$\pm$3.57     &5.1$\pm$2.97\\
005  &   2019-03-20T22:38:35 & 1.066         &        5.43$\pm$2.30     &4.4$\pm$1.94\\
006  &   2019-03-23T22:17:34 & 0.996         &        6.02$\pm$2.46     &4.9$\pm$2.07\\
007  &   2019-03-26T06:32:34 & 1.166         &        5.14$\pm$2.10     &4.2$\pm$1.77\\
008  &   2019-03-26T01:17:35 & 0.302         &        6.62$\pm$4.7      &5.4$\pm$3.87\\
009  &   2019-03-26T01:22:35 & 2.040         &        3.92$\pm$1.38     &3.2$\pm$1.18\\
010  &   2019-04-01T18:40:36 & 1.366         &        2.04$\pm$1.28     &1.7$\pm$1.06\\
011  &   2019-04-04T02:13:36 & 3.264         &        2.15$\pm$0.81     &1.8$\pm$0.69\\
012  &   2019-04-07T14:39:34 & 2.879         &        1.79$\pm$0.86     &1.5$\pm$0.72\\
013  &   2019-04-10T06:20:35 & 2.977         &        1.68$\pm$7.51     &1.4$\pm$0.63\\
014  &   2019-04-16T12:23:04 & 2.814         &        1.20$\pm$0.72     &1.0$\pm$0.60\\
015  &   2019-04-19T00:53:36 & 2.974         &        2.02$\pm$0.82     &1.7$\pm$0.69\\
016  &   2019-04-22T08:25:34 & 3.179         &        3.08$\pm$1.00     &2.5$\pm$0.85\\
017  &   2019-04-25T06:38:34 & 2.972         &        4.38$\pm$1.21     &3.6$\pm$1.06\\
018  &   2019-04-27T07:56:35 & 0.092         &        $\leq$10.82       &$\leq$8.9\\
020  &   2019-05-03T02:39:35 & 3.014         &        6.16$\pm$1.45     &5.1$\pm$1.29\\
021  &   2019-05-06T03:58:34 & 2.829         &        3.89$\pm$1.17     &3.2$\pm$1.01\\
022  &   2019-05-08T10:00:35 & 2.707         &        7.31$\pm$1.65     &6.0$\pm$1.48\\
024  &   2019-05-15T07:44:34 & 2.505         &        5.02$\pm$1.44     &4.1$\pm$1.25\\  
025  &   2019-05-18T07:27:36 & 2.807         &        2.42$\pm$0.94     &2.0$\pm$0.8\\
026  &   2019-05-21T10:21:36 & 2.667         &        2.98$\pm$1.06     &2.4$\pm$0.9\\
027  &   2019-05-24T00:32:36 & 2.829         &        4.17$\pm$1.23     &3.4$\pm$1.06\\
029  &   2019-05-27T08:14:36 & 1.493         &        2.01$\pm$1.16     &1.6$\pm$0.96\\
030  &   2019-05-31T04:42:36 & 1.536         &        2.60$\pm$1.30     &2.1$\pm$1.09\\
031  &   2019-06-04T06:21:36 & 2.322         &        2.09$\pm$0.96     &1.7$\pm$0.81\\
032  &   2019-10-11T22:12:36 & 1.618         &        74.74$\pm$0.68    &61.2$\pm$8.27\\
033  &   2019-10-17T20:27:35 & 1.094         &        96.33$\pm$0.94    &78.9$\pm$11.01\\
034  &   2019-10-23T21:11:35 & 1.656         &        95.87$\pm$0.76    &78.6$\pm$10.02\\
035  &   2019-10-26T16:02:34 & 2.008         &        163.8$\pm$0.9     &134.2$\pm$15.29\\
036  &   2019-10-31T04:45:36 & 1.930         &        66.36$\pm$0.59    &54.4$\pm$7.26\\
037  &   2019-11-03T14:01:14 & 0.499         &        130.9$\pm$0.2.    &107.3$\pm$17.08\\
038  &   2019-11-07T15:08:35 & 1.878         &        32.03$\pm$0.42    &26.2$\pm$4.30\\
039  &   2019-11-09T14:42:36 & 2.030         &        48.43$\pm$0.49    &39.7$\pm$5.65\\
040  &   2019-11-12T16:25:36 & 1.905         &        95.07$\pm$0.71    &77.9$\pm$9.68\\
\hline  
ObsID      &  ObsDate   &      Exposure       &      Count rate       &  Count rate   &Flux   &Flux             \\ 
 11186-    &                 &      total              &      0.3-2 keV       &  2-10 keV    &0.3-2 keV&                2-10 keV \\ 
 stacking &                 &       (ks)            &       (10$^{-3}$ cts/s)  &(10$^{-3}$ cts/s)&(10$^{-13}$erg/s/ cm$^{2}$) & ($10^{-13}$erg/s/cm$^{2}$)           \\ 
\hline 
001-009 & March  &8.783&3.94$\pm$0.67&0.24$\pm$0.20&2.71$^{-0.66}_{+0.52}$         &   0.19$^{-0.18}_{+0.19}$  \\
010-018 & April  &22.52&2.41$\pm$0.34&0.39$\pm$0.15&1.66$^{-0.37}_{+0.27}$         &   0.32$^{-0.18}_{+0.20}$ \\
020-031 & May-Jun&24.71&4.13$\pm$0.42&0.19$\pm$0.11&2.84$^{-0.57}_{+0.38}$          &   0.15$^{-0.11}_{+0.12}$\\
001-031 & Mar-Jun&56.01&3.40$\pm$0.25&0.309$\pm$0.09&2.34$^{-0.37}_{+0.10}$            &  0.25$^{-0.08}_{+0.10}$\\
032-040&Oct-Nov  &14.62&85.29$\pm$0.24&$\textless$0.1041&58.70$^{-10.24}_{+4.99}$     & $\textless$0.084\\
\hline  
\end{tabular*}  
\end{table}

\begin{table}
\caption{Stacking X-ray spectral fitting results. The uncertainties are given at 90\% confidence level. \\NOTE: The Galactic absorbed models ''phabs * (1. or 2. or 3. or 4. or 5. ) '' are used for fitting (1. powerlaw,  2. powerlaw + powerlaw,    3.  zbbody,   4. zbbody+powerlaw, 5. zbbody+zbbody). \\
The best-fit model is 4 for the stacking O01\_31,  and 5 for the stacking O32\_40.\\}	\label{tab:2}
\begin{tabular*}{16cm}{p{3cm}p{3cm}p{3cm}p{3cm}p{3cm}}  
\hline
Stacking O01\_31 & & & &\\
            &Model & Parameter &  Value      &     C-Statistic/d.o.f \\ 
\hline
1    &  powerlaw    & $\Gamma$  & $4.95 ^{+0.44}_{-0.40}$      &  64.4/36 \\
2    &  powerlaw1  & $\Gamma_1$  &$6.10 ^{+0.90}_{-0.73}$    &            \\
      & powerlaw2  & $\Gamma_2$   &$1.34 ^{+0.77}_{-0.89}$     &   33.4/34\\
3    & zbbody     & $kT$ (keV)    & $0.077 ^{+0.008}_{-0.008}$  &  104.72/36\\
4    & zbbody     & $kT$ (keV)    & $0.056 ^{+0.009}_{-0.009}$  & \\
     & powerlaw   & PhoIndex  &  $1.95 ^{+0.64}_{-0.59}$     &   33.6/34\\
\hline  
Stacking O32\_40 & & &\\
& Model &  Parameter   & Value   &                     C-Statistic/d.o.f \\ 
\hline
1    &  powerlaw  & $\Gamma$ & $6.75 ^{+0.20}_{-0.19}$    &     63.6/46\\
3   &  zbbody    & $kT$ (keV)   &  $0.060 ^{+0.002}_{-0.002}$   &   98.2/46\\
4    & zbbody     & $kT$ (keV)   &  $0.053 ^{+0.006}_{-0.004}$   &\\
     & powerlaw   & PhoIndex  &  $6.00 ^{+0.70}_{-0.96}$    &    46.4/44\\
5    & zbbody1    & $kT_1$ (keV)    & $0.051 ^{+0.003}_{-0.004}$   & \\
     & zbbody2    & $kT_2$ (keV)    & $0.120 ^{+0.028}_{-0.020}$   &   43.9/44\\
\end{tabular*}  
\end{table} 

\begin{center} 
\begin{table}
\caption{Host-subtracted photometry\\}	\label{tab:host_subtracted}
\begin{tabular*}{18cm}{p{3cm}p{3cm}p{3cm}p{3cm}p{3cm}p{3cm}}
\hline
MJD&ASASSN-g&MJD&ASASSN-v&MJD&ZTF-g\\
day&mag&day&mag&day&mag\\
\hline
58515.25	&18.84	$\pm$0.3	&58529.44	&17.7	$\pm$0.29	&58756.47	&18.04	$\pm$0.08	\\
58523.22	&18.45	$\pm$0.38	&58530.21	&16.84	$\pm$0.17	&58745.52	&18.15	$\pm$0.09	\\
58527.89	&18.33	$\pm$0.27	&58534.26	&16.76	$\pm$0.1	&58737.51	&17.89	$\pm$0.08	\\
58529.54	&17.18	$\pm$0.13	&58535.24	&16.58	$\pm$0.1	&58736.49	&17.94	$\pm$0.07	\\
58534.26	&16.64	$\pm$0.09	&58536.3	&16.08	$\pm$0.04	&58733.49	&17.79	$\pm$0.08	\\
58535.12	&16.31	$\pm$0.07	&58537.47	&16.18	$\pm$0.04	&58609.18	&16.1	$\pm$0.06	\\
58536.33	&16.13	$\pm$0.03	&58538.4	&16.01	$\pm$0.03	&58609.18	&16.12	$\pm$0.06	\\
58537.47	&16.1	$\pm$0.03	&58539.34	&16.02	$\pm$0.08	&58606.17	&16.09	$\pm$0.06	\\
58538.69	&15.9	$\pm$0.03	&58540.27	&15.95	$\pm$0.04	&58606.15	&16.1	$\pm$0.06	\\
58539.26	&15.99	$\pm$0.05	&58541.52	&15.88	$\pm$0.03	&58602.19	&15.43	$\pm$0.06	\\
58540.27	&15.89	$\pm$0.03	&58542.88	&15.8	$\pm$0.03	&58602.19	&16	$\pm$0.05	\\
58541.57	&15.82	$\pm$0.02	&58543.3	&15.83	$\pm$0.03	&58598.21	&15.91	$\pm$0.06	\\
58542.88	&15.75	$\pm$0.02	&58544.37	&15.87	$\pm$0.03	&58598.19	&15.9	$\pm$0.06	\\
58543.4	&15.79	$\pm$0.02	&58545.32	&15.86	$\pm$0.04	&58594.19	&15.82	$\pm$0.06	\\
58544.37	&15.82	$\pm$0.02	&58546.02	&15.89	$\pm$0.04	&58594.19	&15.78	$\pm$0.06	\\
58545.93	&15.68	$\pm$0.02	&58547.58	&15.89	$\pm$0.03	&58591.15	&15.74	$\pm$0.06	\\
58546.02	&15.84	$\pm$0.03	&58548.21	&15.8	$\pm$0.04	&58591.15	&15.76	$\pm$0.06	\\
58547.87	&15.82	$\pm$0.02	&58549.15	&15.78	$\pm$0.03	&58588.17	&15.66	$\pm$0.06	\\
58548.21	&15.75	$\pm$0.04	&58550.31	&15.5	$\pm$0.08	&58588.17	&15.67	$\pm$0.06	\\
58549.28	&15.78	$\pm$0.03	&58551.13	&15.8	$\pm$0.04	&58584.21	&15.56	$\pm$0.05	\\
58550.31	&15.65	$\pm$0.06	&58552.01	&15.72	$\pm$0.04	&58584.21	&15.56	$\pm$0.06	\\
58551.31	&15.76	$\pm$0.03	&58554.27	&15.47	$\pm$0.03	&58581.21	&15.52	$\pm$0.06	\\
58554.39	&15.41	$\pm$0.01	&58555.21	&15.44	$\pm$0.03	&58581.17	&15.52	$\pm$0.06	\\
58555.21	&15.4	$\pm$0.01	&58556.18	&15.49	$\pm$0.03	&58575.2	&15.41	$\pm$0.06	\\
58557.25	&15.3	$\pm$0.04	&58557.25	&15.33	$\pm$0.04	&58575.2	&15.41	$\pm$0.06	\\
58560.85	&15.52	$\pm$0.06	&58560.86	&15.56	$\pm$0.06	&58572.19	&15.34	$\pm$0.06	\\
58561.28	&15.15	$\pm$0.03	&58561.28	&15.18	$\pm$0.04	&58572.19	&15.33	$\pm$0.06	\\
58563.88	&15.16	$\pm$0.03	&58563.88	&15.19	$\pm$0.04	&58567.19	&15.29	$\pm$0.06	\\
58564.9	&15.34	$\pm$0.03	&58564.9	&15.38	$\pm$0.04	&58567.17	&15.29	$\pm$0.05	\\
58565.91	&15.22	$\pm$0.08	&58565.91	&15.25	$\pm$0.08	&58561.23	&15.28	$\pm$0.05	\\
58566.63	&15.32	$\pm$0.02	&58566.63	&15.36	$\pm$0.03	&58556.27	&15.29	$\pm$0.06	\\
58567.8	&15.39	$\pm$0.01	&58567.47	&15.4	$\pm$0.03	&58547.29	&15.71	$\pm$0.06	\\
58568.28	&15.43	$\pm$0.05	&58568.24	&15.42	$\pm$0.03	&58547.29	&15.75	$\pm$0.06	\\
58570.14	&15.38	$\pm$0.03	&58569.39	&15.4	$\pm$0.03	&58538.26	&15.83	$\pm$0.06	\\
58571.19	&15.39	$\pm$0.02	&58570.24	&15.43	$\pm$0.03	&58538.25	&15.85	$\pm$0.06	\\
58574.24	&15.52	$\pm$0.02	&58571.19	&15.43	$\pm$0.03	&58534.24	&16.43	$\pm$0.06	\\
58575.23	&15.46	$\pm$0.02	&58572.13	&15.37	$\pm$0.04	&	&		\\
58577.09	&15.54	$\pm$0.02	&58573.16	&15.42	$\pm$0.03	&	&		\\
58578.85	&15.53	$\pm$0.02	&58574.15	&15.59	$\pm$0.03	&	&		\\
58580.2	&15.6	$\pm$0.02	&58575.23	&15.5	$\pm$0.03	&	&		\\
58583.19	&15.65	$\pm$0.02	&58577.09	&15.58	$\pm$0.03	&	&		\\
58593.16	&16.03	$\pm$0.05	&58578.64	&15.56	$\pm$0.03	&	&		\\
58594.04	&15.77	$\pm$0.05	&58580.25	&15.63	$\pm$0.03	&	&		\\
58596.05	&15.87	$\pm$0.04	&58581.35	&15.59	$\pm$0.03	&	&		\\
58600.4	&15.98	$\pm$0.02	&58582.28	&15.62	$\pm$0.03	&	&		\\
58601.04	&16.04	$\pm$0.04	&58583.19	&15.7	$\pm$0.03	&	&		\\
58602.18	&16.06	$\pm$0.23	&58589.29	&15.75	$\pm$0.04	&	&		\\
58603.09	&16.08	$\pm$0.04	&58591.18	&15.82	$\pm$0.05	&	&		\\
58607.04	&16.49	$\pm$0.12	&58592.27	&15.9	$\pm$0.05	&	&		\\
58608.11	&16.03	$\pm$0.03	&58593.16	&16.1	$\pm$0.06	&	&		\\
\hline
\end{tabular*}  
\end{table} 
\end{center} 

\begin{center} 
\begin{table}
\caption{continued\\}
\begin{tabular*}{19cm}{p{3cm}p{3cm}p{3cm}p{3cm}p{3cm}p{3cm}}
\hline
MJD&ASASSN-g&MJD&ASASSN-v&MJD&ZTF-g\\
day&mag&day&mag&day&mag\\
\hline
58611.12	&16.25	$\pm$0.04	&58594.06	&15.8	$\pm$0.05	&	&		\\
58612.04	&16.28	$\pm$0.05	&58595.19	&15.99	$\pm$0.05	&	&		\\
58616.11	&16.12	$\pm$0.08	&58596.05	&15.92	$\pm$0.05	&	&		\\
58617.99	&16.09	$\pm$0.07	&58597.24	&16.01	$\pm$0.03	&	&		\\
58620.11	&16.81	$\pm$0.21	&58598.28	&16.02	$\pm$0.03	&	&		\\
58621.55	&15.8	$\pm$0.06	&58600.4	&16.04	$\pm$0.03	&	&		\\
58622.46	&16.22	$\pm$0.15	&58601.04	&16.11	$\pm$0.04	&	&		\\
58623.97	&16.11	$\pm$0.04	&58602.18	&15.41	$\pm$0.17	&	&		\\
58625.71	&16.28	$\pm$0.04	&58603.09	&16.15	$\pm$0.05	&	&		\\
58626.96	&16.39	$\pm$0.05	&58605.26	&16.21	$\pm$0.04	&	&		\\
58634	&16.53	$\pm$0.08	&58607.14	&16.48	$\pm$0.07	&	&		\\
58636.7	&16.41	$\pm$0.06	&58608.11	&16.1	$\pm$0.04	&	&		\\
58637.96	&16.66	$\pm$0.08	&58610.25	&16.17	$\pm$0.04	&	&		\\
58766.44	&18.6	$\pm$0.25	&58611.12	&16.34	$\pm$0.05	&	&		\\
58775.47	&17.91	$\pm$0.27	&58612.04	&16.37	$\pm$0.06	&	&		\\
58785.43	&17.71	$\pm$0.16	&58616.19	&16.27	$\pm$0.05	&	&		\\
58790.39	&18.84	$\pm$0.83	&58617.99	&16.16	$\pm$0.07	&	&		\\
58791.27	&18.76	$\pm$0.35	&58618.02	&16.1	$\pm$0.09	&	&		\\
58792.3	&18.87	$\pm$0.34	&58619.99	&17.7	$\pm$0.55	&	&		\\
58795.32	&18.56	$\pm$0.36	&58620.11	&16.96	$\pm$0.24	&	&		\\
58802.43	&18.36	$\pm$0.49	&58621.55	&15.85	$\pm$0.07	&	&		\\
58808.27	&18.71	$\pm$0.29	&58622.46	&16.3	$\pm$0.16	&	&		\\
58813.39	&16.71	$\pm$0.27	&58623.98	&16.35	$\pm$0.05	&	&		\\
58818.06	&18.15	$\pm$0.61	&58625.83	&16.43	$\pm$0.04	&	&		\\
58820.35	&17.88	$\pm$0.13	&58626.97	&16.53	$\pm$0.05	&	&		\\
58821.39	&18.24	$\pm$0.4	&58633.99	&16.55	$\pm$0.06	&	&		\\
58824.39	&18.05	$\pm$0.18	&58636.7	&16.5	$\pm$0.07	&	&		\\
58828.26	&17.09	$\pm$0.26	&58637.96	&16.79	$\pm$0.1	&	&		\\
58830.31	&17.38	$\pm$0.25	&58752.62	&18.15	$\pm$0.18	&	&		\\
58837.42	&18.37	$\pm$0.21	&58755.6	&18.38	$\pm$0.17	&	&		\\
58839.06	&18.86	$\pm$0.42	&58760.6	&18.68	$\pm$0.44	&	&		\\
58849.18	&17.89	$\pm$0.22	&58762.58	&18.25	$\pm$0.14	&	&		\\
58850.2	&17.74	$\pm$0.17	&58763.59	&18.35	$\pm$0.16	&	&		\\
58851.98	&17.99	$\pm$0.4	&58764.55	&18.85	$\pm$0.26	&	&		\\
58866.45	&18.36	$\pm$0.18	&58765.59	&18.51	$\pm$0.18	&	&		\\
58867.44	&18.84	$\pm$0.35	&58771.54	&18.36	$\pm$0.32	&	&		\\
58868.14	&18.06	$\pm$0.3	&58775.47	&18.36	$\pm$0.42	&	&		\\
58869.27	&18.4	$\pm$0.22	&58780.6	&18.66	$\pm$0.27	&	&		\\
58870.14	&18.07	$\pm$0.28	&58785.43	&18.07	$\pm$0.23	&	&		\\
58872.4	&18.05	$\pm$0.12	&58799.58	&18.74	$\pm$0.57	&	&		\\
58874.25	&18.4	$\pm$0.17	&58813.39	&16.84	$\pm$0.31	&	&		\\
58875.28	&18.79	$\pm$0.3	&58818.06	&18.76	$\pm$1.07	&	&		\\
58876.32	&18.1	$\pm$0.43	&58820.35	&18.31	$\pm$0.2	&	&		\\
58878.25	&18.75	$\pm$0.21	&58824.44	&18.5	$\pm$0.2	&	&		\\
58880.26	&18.89	$\pm$0.24	&58825.49	&17.9	$\pm$0.18	&	&		\\
58881.23	&18.35	$\pm$0.22	&58828.37	&17.38	$\pm$0.19	&	&		\\
58882.34	&18.32	$\pm$0.21	&58830.31	&17.64	$\pm$0.32	&	&		\\
58892.38	&18.64	$\pm$0.35	&58836.55	&18.68	$\pm$0.49	&	&		\\
58894.37	&18.35	$\pm$0.23	&58837.53	&18.58	$\pm$0.22	&	&		\\
58895.38	&18.16	$\pm$0.26	&58841.34	&18.87	$\pm$0.26	&	&		\\
\hline
\end{tabular*}  
\end{table} 
\end{center} 

\begin{center} 
\begin{table}
\caption{continued\\}
\begin{tabular*}{19cm}{p{3cm}p{3cm}p{3cm}p{3cm}p{3cm}p{3cm}}
\hline
MJD&ASASSN-g&MJD&ASASSN-v&MJD&ZTF-g\\
day&mag&day&mag&day&mag\\
\hline
58896.38	&15.57	$\pm$0.03	&58850.36	&18.53	$\pm$0.19	&	&		\\
58907.54	&17.57	$\pm$0.19	&58851.98	&18.49	$\pm$0.64	&	&		\\
58920.2	&18.37	$\pm$0.42	&58868.14	&18.59	$\pm$0.5	&	&		\\
58922.91	&17.89	$\pm$0.37	&58870.14	&18.89	$\pm$0.45	&	&		\\
58924.12	&18.76	$\pm$0.59	&58872.4	&18.59	$\pm$0.19	&	&		\\
58928.29	&18.55	$\pm$0.31	&58876.32	&18.66	$\pm$0.73	&	&		\\
58932.42	&18.6	$\pm$0.33	&58895.38	&18.77	$\pm$0.47	&	&		\\
58967.16	&18.6	$\pm$0.31	&58896.38	&15.61	$\pm$0.04	&	&		\\
58976.13	&18.57	$\pm$0.6	&58907.54	&17.88	$\pm$0.25	&	&		\\
58978.12	&18.57	$\pm$0.54	&58922.91	&18.33	$\pm$0.56	&	&		\\
58979.14	&17.11	$\pm$0.11	&58947.32	&17.97	$\pm$0.26	&	&		\\
58990.73	&17.7	$\pm$0.19	&58979.14	&17.31	$\pm$0.14	&	&		\\
58993.72	&18.02	$\pm$0.22	&58990.73	&18.05	$\pm$0.26	&	&		\\
	&		&58993.72	&18.54	$\pm$0.35	&	&		\\
\hline
MJD&ZTF-r&MJD&UVOT-uvw1&MJD&UVOT-uvw2\\
day&mag&day&mag&day&mag\\
\hline
58617.21	&16.37	$\pm$0.05	&58544.86	&14.91	$\pm$0.04	&58544.87	&14.62	$\pm$0.04	\\
58617.21	&16.32	$\pm$0.05	&58553.45	&14.82	$\pm$0.04	&58553.46	&14.6	$\pm$0.04	\\
58609.16	&16.22	$\pm$0.05	&58557.1	&14.7	$\pm$0.04	&58556.11	&14.37	$\pm$0.04	\\
58609.16	&16.25	$\pm$0.05	&58562.95	&14.67	$\pm$0.04	&58562.96	&14.61	$\pm$0.04	\\
58606.17	&16.18	$\pm$0.05	&58565.94	&14.72	$\pm$0.04	&58565.94	&14.57	$\pm$0.04	\\
58602.16	&15.82	$\pm$0.06	&58568.48	&14.88	$\pm$0.04	&58568.77	&14.6	$\pm$0.04	\\
58598.15	&15.96	$\pm$0.05	&58575.06	&14.81	$\pm$0.04	&58568.48	&14.82	$\pm$0.04	\\
58591.14	&15.77	$\pm$0.05	&58577.49	&14.83	$\pm$0.04	&58575.06	&14.68	$\pm$0.04	\\
58591.14	&15.79	$\pm$0.05	&58580.73	&14.9	$\pm$0.04	&58577.49	&14.72	$\pm$0.04	\\
58584.16	&15.65	$\pm$0.05	&58583.39	&14.94	$\pm$0.04	&58580.73	&14.73	$\pm$0.04	\\
58584.16	&15.63	$\pm$0.05	&58589.82	&15.08	$\pm$0.04	&58583.39	&14.75	$\pm$0.04	\\
58582.22	&15.56	$\pm$0.05	&58592.13	&15.12	$\pm$0.04	&58589.82	&14.87	$\pm$0.04	\\
58581.16	&15.55	$\pm$0.05	&58595.96	&15.19	$\pm$0.04	&58592.14	&14.87	$\pm$0.04	\\
58581.15	&15.66	$\pm$0.06	&58598.38	&15.28	$\pm$0.04	&58595.97	&14.95	$\pm$0.04	\\
58575.15	&15.54	$\pm$0.05	&58606.41	&15.41	$\pm$0.04	&58598.38	&15.06	$\pm$0.04	\\
58575.15	&15.56	$\pm$0.05	&58610.16	&15.44	$\pm$0.04	&58600.34	&15.33	$\pm$0.07	\\
58572.15	&15.48	$\pm$0.05	&58612.03	&15.44	$\pm$0.04	&58606.42	&15.1	$\pm$0.04	\\
58567.15	&15.61	$\pm$0.06	&58618.53	&15.56	$\pm$0.04	&58610.17	&15.11	$\pm$0.04	\\
58567.15	&15.65	$\pm$0.06	&58622.12	&15.61	$\pm$0.04	&58612.04	&15.13	$\pm$0.04	\\
58561.18	&15.43	$\pm$0.05	&58625.04	&15.61	$\pm$0.04	&58618.54	&15.22	$\pm$0.04	\\
58556.19	&15.69	$\pm$0.05	&58627.43	&15.68	$\pm$0.04	&58622.12	&15.29	$\pm$0.04	\\
58556.19	&15.53	$\pm$0.05	&58630.45	&15.69	$\pm$0.04	&58625.04	&15.28	$\pm$0.04	\\
58547.17	&15.92	$\pm$0.05	&58634.3	&15.78	$\pm$0.04	&58627.43	&15.33	$\pm$0.04	\\
58547.17	&15.96	$\pm$0.05	&58638.56	&15.85	$\pm$0.04	&58630.43	&15.36	$\pm$0.04	\\
58543.27	&15.95	$\pm$0.05	&58767.93	&17.52	$\pm$0.05	&58634.31	&15.43	$\pm$0.04	\\
58541.29	&16.01	$\pm$0.06	&58774.03	&17.61	$\pm$0.05	&58634.31	&15.43	$\pm$0.04	\\
58540.22	&16.01	$\pm$0.06	&58779.91	&17.68	$\pm$0.05	&58638.56	&15.58	$\pm$0.04	\\
58534.32	&16.82	$\pm$0.06	&58782.99	&17.72	$\pm$0.05	&58767.94	&17.42	$\pm$0.05	\\
58534.32	&16.52	$\pm$0.06	&58788.29	&17.82	$\pm$0.06	&58774.02	&17.55	$\pm$0.05	\\
	&		&58790.59	&17.82	$\pm$0.06	&58779.9	&17.62	$\pm$0.05	\\
	&		&58794.74	&17.98	$\pm$0.06	&58782.99	&17.59	$\pm$0.05	\\
	&		&58796.73	&18.06	$\pm$0.06	&58788.29	&17.71	$\pm$0.05	\\
\hline
\end{tabular*}  
\end{table} 
\end{center} 

\begin{center} 
\begin{table}
\caption{continued\\}
\begin{tabular*}{19cm}{p{3cm}p{3cm}p{3cm}p{3cm}p{3cm}p{3cm}}
\hline
MJD&ZTF-r&MJD&UVOT-uvw1&MJD&UVOT-uvw2\\
day&mag&day&mag&day&mag\\
\hline
	&		&58798.77	&18.1	$\pm$0.06	&58790.6	&17.6	$\pm$0.07	\\
	&		&58799.78	&17.83	$\pm$0.06	&58794.74	&17.74	$\pm$0.05	\\
	&		&58802.99	&17.81	$\pm$0.05	&58796.74	&17.89	$\pm$0.05	\\
	&		&58810.77	&18.13	$\pm$0.07	&58799.78	&17.73	$\pm$0.05	\\
	&		&58818.83	&17.97	$\pm$0.06	&58803	&17.73	$\pm$0.05	\\
	&		&58823.01	&18	$\pm$0.05	&58810.38	&17.83	$\pm$0.06	\\
	&		&58826.93	&18.26	$\pm$0.06	&58818.83	&17.95	$\pm$0.05	\\
	&		&58830.41	&18.04	$\pm$0.05	&58823.02	&17.99	$\pm$0.05	\\
	&		&58840.87	&18.17	$\pm$0.05	&58826.9	&18.02	$\pm$0.08	\\
	&		&58843.19	&18.18	$\pm$0.06	&58830.42	&18.03	$\pm$0.05	\\
	&		&58847.82	&18.25	$\pm$0.05	&58840.87	&18.1	$\pm$0.05	\\
	&		&58851.2	&18.25	$\pm$0.06	&58843.19	&18.09	$\pm$0.05	\\
	&		&58906.65	&18.27	$\pm$0.08	&58847.82	&18.21	$\pm$0.05	\\
	&		&58962.78	&18.4	$\pm$0.06	&58851.2	&18.17	$\pm$0.05	\\
	&		&58967.33	&18.53	$\pm$0.07	&58906.65	&18.29	$\pm$0.08	\\
	&		&58972.18	&18.29	$\pm$0.1	&58911.89	&18.53	$\pm$0.13	\\
	&		&58977.69	&18.48	$\pm$0.06	&58962.78	&18.52	$\pm$0.06	\\
	&		&58982.66	&18.46	$\pm$0.07	&58967.34	&18.44	$\pm$0.06	\\
	&		&	&		&58972.18	&18.48	$\pm$0.09	\\
	&		&	&		&58977.59	&18.28	$\pm$0.05	\\
	&		&	&		&58982.67	&18.4	$\pm$0.06	\\
\hline
MJD&UVOT-uvm2&MJD&UVOT-U&MJD&UVOT-B\\
day&mag&day&mag&day&mag\\
\hline
58544.87	&14.82	$\pm$0.04	&58544.86	&15.37	$\pm$0.06	&58544.86	&15.58	$\pm$0.02	\\
58553.46	&14.81	$\pm$0.04	&58553.45	&15.19	$\pm$0.06	&58553.45	&15.34	$\pm$0.02	\\
58556.11	&14.58	$\pm$0.04	&58556.11	&15.03	$\pm$0.06	&58556.11	&15.26	$\pm$0.03	\\
58562.96	&14.59	$\pm$0.04	&58562.95	&15.12	$\pm$0.06	&58562.95	&15.19	$\pm$0.02	\\
58565.94	&14.59	$\pm$0.04	&58565.94	&15.14	$\pm$0.06	&58565.94	&15.34	$\pm$0.02	\\
58568.77	&14.8	$\pm$0.04	&58568.76	&15.16	$\pm$0.05	&58568.76	&15.36	$\pm$0.02	\\
58568.48	&14.76	$\pm$0.04	&58568.48	&15.16	$\pm$0.06	&58568.48	&15.32	$\pm$0.02	\\
58575.06	&14.77	$\pm$0.04	&58575.06	&15.22	$\pm$0.06	&58575.06	&15.46	$\pm$0.02	\\
58577.49	&14.84	$\pm$0.04	&58577.49	&15.2	$\pm$0.05	&58577.49	&15.47	$\pm$0.02	\\
58580.73	&14.87	$\pm$0.04	&58580.73	&15.3	$\pm$0.05	&58580.73	&15.58	$\pm$0.02	\\
58583.39	&15.02	$\pm$0.04	&58583.39	&15.33	$\pm$0.05	&58583.39	&15.74	$\pm$0.02	\\
58589.82	&15.04	$\pm$0.04	&58589.82	&15.48	$\pm$0.06	&58589.82	&15.73	$\pm$0.02	\\
58592.14	&15.12	$\pm$0.04	&58592.13	&15.49	$\pm$0.05	&58592.13	&15.82	$\pm$0.02	\\
58595.97	&15.26	$\pm$0.04	&58595.96	&15.57	$\pm$0.05	&58595.96	&15.86	$\pm$0.02	\\
58598.38	&15.71	$\pm$0.06	&58598.38	&15.68	$\pm$0.06	&58598.38	&16.21	$\pm$0.02	\\
58600.34	&15.34	$\pm$0.04	&58606.41	&15.88	$\pm$0.06	&58606.41	&16.07	$\pm$0.02	\\
58606.42	&15.38	$\pm$0.04	&58610.16	&15.89	$\pm$0.06	&58610.16	&16.09	$\pm$0.02	\\
58610.17	&15.38	$\pm$0.04	&58612.04	&15.88	$\pm$0.06	&58612.04	&16.36	$\pm$0.02	\\
58612.04	&15.51	$\pm$0.04	&58618.53	&15.98	$\pm$0.06	&58618.53	&16.21	$\pm$0.02	\\
58618.54	&15.58	$\pm$0.04	&58622.12	&15.99	$\pm$0.06	&58622.12	&16.5	$\pm$0.02	\\
58622.12	&15.58	$\pm$0.04	&58625.04	&16.13	$\pm$0.06	&58625.04	&16.36	$\pm$0.02	\\
58625.04	&15.63	$\pm$0.04	&58627.43	&16.11	$\pm$0.06	&58627.43	&16.35	$\pm$0.02	\\
58627.43	&15.64	$\pm$0.04	&58630.42	&16.17	$\pm$0.06	&58630.42	&16.54	$\pm$0.02	\\
\hline
\end{tabular*}  
\end{table} 
\end{center} 

\begin{center} 
\begin{table}
\caption{continued\\}
\begin{tabular*}{19cm}{p{3cm}p{3cm}p{3cm}p{3cm}p{3cm}p{3cm}}
\hline
MJD&UVOT-uvm2&MJD&UVOT-U&MJD&UVOT-B\\
day&mag&day&mag&day&mag\\
\hline
58630.43	&15.7	$\pm$0.04	&58634.3	&16.23	$\pm$0.06	&58634.3	&16.54	$\pm$0.02	\\
58634.31	&15.82	$\pm$0.04	&58638.56	&16.28	$\pm$0.06	&	&		\\
58634.31	&17.87	$\pm$0.05	&	&		&	&		\\
58638.56	&17.96	$\pm$0.06	&	&		&	&		\\
58767.94	&18	$\pm$0.05	&	&		&	&		\\
58774.02	&18.14	$\pm$0.05	&	&		&	&		\\
58779.9	&18.12	$\pm$0.05	&	&		&	&		\\
58782.99	&18.36	$\pm$0.05	&	&		&	&		\\
58788.29	&18.45	$\pm$0.05	&	&		&	&		\\
58790.6	&18.26	$\pm$0.05	&	&		&	&		\\
58794.74	&18.33	$\pm$0.05	&	&		&	&		\\
58796.74	&18.34	$\pm$0.06	&	&		&	&		\\
58799.78	&18.56	$\pm$0.05	&	&		&	&		\\
58803	&18.47	$\pm$0.05	&	&		&	&		\\
58810.38	&18.64	$\pm$0.08	&	&		&	&		\\
58818.83	&18.58	$\pm$0.05	&	&		&	&		\\
58823.02	&18.74	$\pm$0.05	&	&		&	&		\\
58826.9	&18.75	$\pm$0.06	&	&		&	&		\\
58830.42	&18.84	$\pm$0.05	&	&		&	&		\\
58840.87	&18.89	$\pm$0.05	&	&		&	&		\\
\hline
MJD&UVOT-V&MJD&Gaia-G&&\\
day&mag&day&mag&day&mag\\
\hline
58544.87	&15.87	$\pm$0.03	&58586.75	&15.5	\\
58553.46	&15.7	$\pm$0.03	&58621.48	&16.11	\\
58556.11	&15.3	$\pm$0.04	&58621.55	&16.21	\\
58562.96	&15.3	$\pm$0.03	&58636.98	&16.42	\\
58565.94	&15.4	$\pm$0.03	&58637.06	&16.4	\\
58568.48	&15.77	$\pm$0.03	&	&	\\
58575.06	&15.51	$\pm$0.03	&	&	\\
58577.49	&15.76	$\pm$0.03	&	&	\\
58580.73	&15.69	$\pm$0.03	&	&	\\
58583.39	&15.86	$\pm$0.03	&	&	\\
58589.82	&16.05	$\pm$0.03	&	&	\\
58592.14	&15.9	$\pm$0.03	&	&	\\
58595.97	&16.07	$\pm$0.03	&	&	\\
58598.38	&16.13	$\pm$0.03	&	&	\\
58606.42	&16.43	$\pm$0.03	&	&	\\
58610.17	&16.32	$\pm$0.03	&	&	\\
58612.04	&16.26	$\pm$0.03	&	&	\\
58618.54	&16.46	$\pm$0.03	&	&	\\
58622.12	&16.41	$\pm$0.03	&	&	\\
58625.04	&16.82	$\pm$0.03	&	&	\\
58627.43	&16.6	$\pm$0.03	&	&	\\
58630.43	&17.11	$\pm$0.03	&	&	\\
58634.31	&16.62	$\pm$0.03	&	&	\\
58638.56	&16.91	$\pm$0.03	&	&	\\
\hline
\end{tabular*}  
\end{table} 
\end{center}

\end{document}